\documentclass[twocolumn,trackchanges]{aastex62}

\hypersetup{linkcolor=red}
\newcommand{\angstrom}{\textup{\AA}} 
 
\newcommand{\msun}{M_\odot} \newcommand{\rsun}{R_\odot}
\newcommand{\ms}{M_\text s} \newcommand{\rs}{R_\text s}
\newcommand{\mj}{M_\text J} \newcommand{\rj}{R_\text J}
\renewcommand{\mp}{M_\text p} \newcommand{\rp}{R_\text p}
 
\newcommand{\rplam}{R_{\text p, \lambda}} 
\newcommand{\rplamnot}{R_{\text p, \lambda, 0}} 
\newcommand{\rprs}{\rp/\rs} 
\newcommand{\tef}{T_\text{eff}} \newcommand{\teq}{T_\text{eq}}
\newcommand{\bt}{\bm{\theta}} \newcommand{\ba}{\bm{\alpha}}
\newcommand{\Fdiv}{F_\text{div}} \newcommand{\Ftarg}{F_\text{W43}}
\newcommand{\Fcomp}{F_\mathrm{C_n}} \newcommand{\Fdet}{F_\text{det}}
\newcommand{\water}{$\rm H_2O$} 
\newcommand{\halpha}{$\rm H\alpha$}
\newcommand\numberthis{\addtocounter{equation}{1}\tag{\theequation}}
\newcommand{\Tchord}{T_\text{chord}} \newcommand{\Thet}{T_\text{het}}
\newcommand{\fhet}{f_\text{het}} \newcommand{\Shet}{S_{\lambda,\text{het}}}
\newcommand{\Schord}{S_{\lambda,\text{chord}}}

\usepackage{amsmath, bm, hyperref, ulem}
\usepackage{savesym} \savesymbol{tablenum}
\usepackage[detect-weight=true, detect-family=true]{siunitx}
\restoresymbol{SIX}{tablenum}
\usepackage[T1]{fontenc}

\graphicspath{{./}{Figures/}}

\received{June 12, 2019} \revised{October 25, 2019} \accepted{November 7, 2019}
\submitjournal{AJ}

\shorttitle{ACCESS: Visual to NIR Transmission Spectrum of 
WASP-43$\lowercase{\text b}$} \shortauthors{Weaver et al.}

\begin{document} 
\title{ACCESS: A Visual to Near-infrared Spectrum of the Hot Jupiter 
WASP-43b with Evidence of \water, but no evidence of Na or K} 

\correspondingauthor{Ian C. Weaver} \email{iweaver@cfa.harvard.edu}

\author[0000-0001-6205-6315]{Ian C. Weaver}
\affiliation{Center for Astrophysics ${\rm \mid}$ Harvard {\rm \&} Smithsonian, 
60 Garden St, Cambridge, MA 02138, USA}

\author[0000-0003-3204-8183]{Mercedes L\'opez-Morales} 
\affiliation{Center for Astrophysics ${\rm \mid}$ Harvard {\rm \&} Smithsonian, 
60 Garden St, Cambridge, MA 02138, USA}

\author[0000-0001-9513-1449]{N\'estor Espinoza} 
\altaffiliation{Bernoulli Fellow, IAU-Gruber Fellow}
\affiliation{Max-Planck-Institut f\"{u}r Astronomie, K\"{o}nigstuhl  17, 
69117 Heidelberg, Germany}

\author[0000-0002-3627-1676]{Benjamin V. Rackham}
\altaffiliation{51 Pegasi b Fellow}
\affiliation{Department of Earth, Atmospheric and Planetary Sciences, and Kavli Institute for Astrophysics and Space Research, Massachusetts Institute of Technology, Cambridge, MA 02139, USA}
\affiliation{Department of Astronomy/Steward Observatory, 
The University of Arizona, 933 N. Cherry Avenue, Tucson, AZ 85721, USA}
\affiliation{Earths in Other Solar Systems Team, NASA Nexus for Exoplanet System
Science}

\author{David J. Osip}
\affiliation{Las Campanas Observatory, Carnegie Institution of
Washington, Colina el Pino, Casilla 601 La Serena, Chile}

\author{D\'aniel Apai} 
\affiliation{Department of Astronomy/Steward Observatory, The
University of Arizona, 933 N. Cherry Avenue, Tucson, AZ 85721, USA}
\affiliation{Earths in Other Solar Systems Team, NASA Nexus
for Exoplanet System Science}
\affiliation{Lunar and Planetary Laboratory, The University of Arizona, 
1640 E. Univ. Blvd, Tucson, AZ 85721} 

\author[0000-0002-5389-3944]{Andr\'es Jord\'an}
\affiliation{Facultad de Ingenier\'ia y Ciencias, Universidad Adolfo
Ib\'a\~nez, Av.\ Diagonal las Torres 2640, Pe\~nalol\'en, Santiago, Chile}
\affiliation{Millennium Institute for Astrophysics, Av.\ Vicu\~{n}a Mackenna
4860, 782-0436 Macul, Santiago, Chile}

\author[0000-0003-2831-1890]{Alex Bixel} 
\affiliation{Department of Astronomy/Steward Observatory, The
University of Arizona, 933 N. Cherry Avenue, Tucson, AZ 85721, USA}
\affiliation{Earths in Other Solar Systems Team, NASA Nexus for Exoplanet System
Science}

\author[0000-0002-8507-1304]{Nikole K. Lewis}
\affiliation{Department of Astronomy and Carl Sagan Institute, 
Cornell University, 122 Sciences Drive, 14853, Ithaca, NY, USA}

\author[0000-0003-4157-832X]{Munazza K. Alam}
\affiliation{Center for Astrophysics ${\rm \mid}$ Harvard {\rm \&} Smithsonian, 
60 Garden St, Cambridge, MA 02138, USA}

\author[0000-0002-4207-6615]{James Kirk}
\affiliation{Center for Astrophysics ${\rm \mid}$ Harvard {\rm \&} Smithsonian, 
60 Garden St, Cambridge, MA 02138, USA}

\author{Chima McGruder} 
\affiliation{Center for Astrophysics ${\rm \mid}$ Harvard {\rm \&} Smithsonian, 
60 Garden St, Cambridge, MA 02138, USA}

\author[0000-0003-0650-5723]{Florian Rodler} 
\affiliation{European Southern Observatory, Alonso de Cordova
3107, Vitacura, Santiago de Chile}  

\author{Jennifer Fienco} 
\affiliation{Instituto de Astrof\'isica, Facultad de
F\'isica, Pontificia Universidad Cat\'olica de Chile, Av. Vicu\~na Mackenna
4860, 7820436 Macul, Santiago, Chile}


\begin{abstract} 
We present a new ground-based visual transmission spectrum of the hot Jupiter
WASP-43b, obtained as part of the ACCESS Survey. The spectrum was derived from
four transits observed between 2015 and 2018, with combined wavelength coverage
between \SIrange{5300}{9000}{\angstrom} and an average photometric precision of
708 ppm in \SI{230}{\angstrom} bins. We perform an atmospheric retrieval of our
transmission spectrum combined with literature \textit{HST/WFC3} observations to
search for the presence of clouds/hazes as well as Na, K, \halpha, and \water\
planetary absorption and stellar spot contamination over a combined spectral
range of \SIrange{5318}{16420}{\angstrom}. We do not detect a statistically
significant presence of \ion{Na}{1} or \ion{K}{1} alkali lines, or \halpha\ in
the atmosphere of WASP-43b. 
We find that the observed transmission
spectrum can be best explained by a combination of heterogeneities on
the photosphere of the host star and a clear planetary atmosphere with
\water. This model yields a log-evidence of $8.26\pm0.42$ higher
than a flat (featureless) spectrum. In particular, the observations
marginally favor the presence of large, low-contrast spots over the four
ACCESS transit epochs with an average covering fraction
$f_\text{het} = 0.27^{+0.42}_{-0.16}$, and temperature contrast
$\Delta T = \SI{132}{K} \pm \SI{132}{K}$. Within the planet's
atmosphere, we recover a log \water\ volume mixing ratio of
$-2.78^{+1.38}_{-1.47}$, which is
consistent with previous \water\ abundance determinations for this planet.
\end{abstract}

\keywords{planets and satellites: atmospheres --- 
planets and satellites: individual (WASP-43b) --- 
stars: activity --- stars: starspots --- techniques: spectroscopic}


\section{Introduction} \label{sec:intro}
Observations of exoplanetary atmospheres offer the possibility of understanding
the atmospheric physical properties and chemical composition of those worlds, as
well as providing clues to their formation and evolution histories (e.g.
\citealt{oberg13}; \citealt{moses13}; \citealt{mordasini16},
\citealt{espinoza2017}). The first comparison studies of exoplanetary
atmospheres using transmission spectra (see, e.g.,
\citealt{sing2016,crossfield_kreidberg_2017}), found evidence of a gradual
transition between clear and cloudy atmospheres, but no clear correlation of
that transition with other system parameters, such as planetary mass, gravity,
effective temperature, or stellar irradiation levels, and the chemical
composition of the star. Recently \citealt{Pinhas2019} reanalyzed the
\citealt{sing2016} sample and concluded that the majority of hot Jupiters have
atmospheres consistent with sub-solar \water\ abundances, with the log of
those values ranging from $-5.04^{+0.46}_{-0.30}$ to $-3.16^{+0.66}_{-0.69}$.

High altitude clouds/hazes have been inferred in the atmosphere of a number of
exoplanets from scattering slopes in the visible \citep[e.g.][]{pont2008,
pont2013, sing2009, sing2011, sing2013, gibson2013}, and from  the damping of
pressure-broadened alkali \ion{Na}{1} and \ion{K}{1} lines originating from
deeper within the atmosphere \citep[e.g.][]{charbonneau2002,wakeford2014}. The
first detections of exoplanets with potentially clear atmospheres have only
recently been made, e.g. WASP-96b \citep{nikolov2018}.

This field is currently at the point where a number of efforts are underway to
identify what system parameters, if any, correlate with the observed
atmospheric properties of exoplanets. For example, the relationship between
chemical abundance in an exoplanet's atmosphere and planet mass is actively
being explored \citep{helling2016, sing2016, kreidberg2014, fraine2014}.
Ground-based (e.g., ACCESS~\footnote{Arizona-CfA-Cat\'olica-Carnegie Exoplanet
Spectroscopy Survey \citep{rackham2017}}, GPIES~\footnote{Gemini Planet
Imager Extra Solar Survey \citep{nielsen2019}}, VLT FORS2~\footnote{Very
Large Telescope FOcal Reducer and Spectrograph \citep{nikolov2018}},
LRG-BEASTS~\footnote{Low Resolution Ground-Based Exoplanet Atmosphere
Survey using Transmission Spectroscopy \citep{kirk2018}}) and space based
surveys (PanCET~\footnote{Panchromatic Comparative Exoplanetology Treasury
\citep{wakeford2017}}) are working to provide homogeneous spectra of a
statistically significant number of exoplanet atmospheres in the search for
these correlations. In this paper, we present the ground based visual to
near-infrared (NIR) transmission spectrum of the hot Jupiter WASP-43b
obtained as part of the ACCESS survey.

WASP-43b ($\mp=2.052\pm0.053$ $\mj$, $\rp=1.036\pm0.012$
$\rj$, $\teq=1440\substack{+40\\-39}$ K; \citealt{gillon2012}) is a hot Jupiter
discovered by \citet{hellier2011} transiting a $V=12.4$ K7V type dwarf star
($\ms=0.717\pm0.025$ $\msun$, $\rs=0.667\pm0.010$ $\rsun$, $\tef=4520\pm120$ K;
\citealt{gillon2012}) every 0.81 days. WASP-43 is unusually active for its
stellar type, as indicated by the presence of strong Ca H and K lines and
perhaps due to star-planet interactions in this very short period system
\citep{staab2017}. Spitzer secondary eclipse data in the \SI{3.6}{\micron} and
\SI{4.5}{\micron} bands indicate brightness temperatures of $1670\pm23$ K and
$1514\pm25$ K, respectively, which rule out a strong thermal inversion in the
planet's dayside photosphere \citep{blecic2014}. In addition, thermal emission
observed in the $K-$band \citep{chen2014} agrees with atmospheric models of
WASP-43b, which predict poor day-to-night heat redistribution in an atmosphere
with no thermal inversion present \citep{kataria2015}.

The presence of water on the dayside of the planet was observed with
\textit{HST/WFC3} emission measurements by \citet{stevenson2014}, with
additional transmission observations by \citet{kreidberg2014} finding
water abundances comparable to solar values. In the visual regime, the Gran
Telescopio Canarias's (GTC) visual System for Imaging and low Resolution
Integrated Spectroscopy (OSIRIS) instrument shows a tentative excess in 
$\rp/\rs$ at \ion{Na}{1} and a complete lack of one near the \ion{K}{1} doublet
\citep{murgas2014}.  That same study also notes a trend of increasing
planet-to-star radius ratio from \SIrange{6200}{7200}{\angstrom} and decreasing
trend redward of \SI{7200}{\angstrom}. They attribute this pattern to the
possible presence of VO and TiO.

In this work, we search further for the presence of \ion{Na}{1}, \ion{K}{1}, and
\halpha\ with new visual transit observations from \textit{Magellan/IMACS}. In
addition, we combine the \textit{HST/WFC3} transmission spectrum of WASP-43b
\citep{kreidberg2014} with the new visual data to produce the full visual to NIR
spectrum spanning \SIrange{5317.90}{16420}{\angstrom} that can further constrain
the water absorption features present in the infrared spectrum and provide
new information about water abundance. In our analysis, we find that the
atmosphere of WASP-43b is best described by a clear atmosphere with water 
abundance consistent with solar. The planet's spectrum is also contaminated with 
stellar heterogeneity. 

This paper is structured as follows: In Section~\ref{sec:obs}, we present our
\textit{Magellan/IMACS} observations. In Section~\ref{sec:data_red_lc_analysis} 
we outline the data reduction process used in our observations and describe the
selection of wavelength bins to optimize the signal-to-noise ratio (SNR),
search for atomic features, and compare to other results. We present the
detrended white and binned light curves for each dataset. In
Section~\ref{sec:stell_act}, we give a qualitative analysis of the impact of
observational stellar activity on the resulting combined transmission spectra
from different visits. In Section~\ref{sec:tspec}, we present the final
transmission spectrum and also compare to the results of \cite{murgas2014}. In
Section~\ref{sec:bin_kreidberg}, we present the results of a retrieval modeling
analysis on the combined ACCESS and HST transmission spectrum to find the best
fit transmission model when the presence of a heterogeneous stellar photosphere
is also taken into account. We summarize and conclude in
Section~\ref{sec:conclusion}.

\section{Observations} \label{sec:obs} 
\subsection{General setup}
We observed four transits of WASP-43b between 2015 and 2018 with the 6.5 meter
Magellan Baade Telescope and Inamori-Magella Areal Camera and Spectrograph
(IMACS, \citealt{dressler2006}) as part of ACCESS. For this study, we used the
IMACS f/2 camera, which has a \SI{27.4}{\arcmin} diameter field of view (FoV).
With this large FoV, IMACS is able to observe several nearby comparison stars
simultaneously to WASP-43 to effectively remove common instrumental and
atmospheric systematics. We selected comparison stars less than 0.5 magnitude
brighter and 1 magnitude fainter that WASP-43 and closest in $B-V/J-K$ color
space, following \citet{rackham2017}. The selected comparison stars are shown
in Table~\ref{tab:comp_stars}. We used a custom designed multi-slit mask with
$\SI{12}{\arcsecond}\times\SI{20}{\arcsecond}$ slits for the target and
comparison stars. We used a similar calibration mask with
$\SI{0.5}{\arcsecond}\times\SI{20}{\arcsecond}$ for arc lamp wavelength
calibrations. We used a 300 line per mm grating with a $17.5^\circ$ blaze
angle for all four datasets to achieve an average resolving power of
$R\sim1,200$, or approximately \SI{4.7}{\angstrom} per resolution element and
access a full wavelength coverage of \SIrange{4500}{9260}{\angstrom}. In
practice, the SNR redward of \SI{9000}{\angstrom} and blueward of
\SI{5300}{\angstrom} dropped to less than 25\% of peak counts. For this reason,
we omitted measurements outside of this range for the rest of the study. We
omitted data taken at an airmass (Z > 2.0) and/or during twilight as well.
\begin{deluxetable}{ccccccc}[htb]
    \caption{Target and comparison star magnitudes and coordinates from the
	    UCAC4 catalog
	    (\url{http://vizier.u-strasbg.fr/viz-bin/VizieR?-source=I/322A&-to=3}).} 
    \label{tab:comp_stars}
    \tablehead{\colhead{Star} & \colhead{RA} & \colhead{Dec} & \colhead{B}
                              & \colhead{V} & \colhead{J} & \colhead{K}}
    \startdata 
	WASP-43 & 10:19:38.0 & -09:48:22.6 & 13.8 & 12.5 & 10.0 & 9.3 \\
        1       & 10:19:23.6 & -09:36:24.9 & 13.1 & 12.5 & 11.4 & 11.0 \\
        2       & 10:19:30.7 & -09:50:58.2 & 13.3 & 12.7 & 11.6 & 11.3 \\
        3       & 10:20:03.3 & -09:34:16.3 & 13.3 & 12.8 & 11.6 & 11.2 \\
        4       & 10:18:55.5 & -09:51:00.4 & 13.9 & 13.0 & 11.5 & 11.0 \\
        5       & 10:19:37.8 & -09:32:22.0 & 14.0 & 13.2 & 11.6 & 11.2 \\
        6       & 10:19:33.5 & -09:41:45.9 & 14.5 & 13.3 & 10.8 & 10.1
    \enddata 
\end{deluxetable}

\subsection{Data Collection}
We collected the two 2015 datasets on 14 Feb 2015 (UT 01:03 -- 08:55, 433
science images) and 09 Mar 2015 (UT 04:35 -- 08:22, 119 science images),
collected a 2017 dataset on 10 Apr 2017 (UT 00:35 -- 03:35, 197 science images),
and a final 2018 dataset on 03 Jun 2018 (UT 23:36 -- 02:29, 156 science images).
During the 2017 and 2018 nights of observing, we introduced a blocking filter to
reduce contamination from light at higher orders while also truncating the
spectral range to \SIrange{5300}{9200}{\angstrom}. We made observations in
Multi-Object Spectroscopy mode with $2{\times}2$ binning in TURBO readout mode
(30\,s) for the 2015 datasets and in $2{\times}2$ binning in Fast readout mode
(31\,s) for the 2017 and 2018 datasets to take advantage of the reduced readout
noise. During the observations, we adjusted the individual exposure times
between 20--60 seconds to keep the number of counts per pixel roughly between
30,000--35,000 counts (ADU; gain = 1e$^-$/ADU on f/2 camera), i.e., within the
linearity limit of the CCD \citep{bixel2019}.

With the calibration mask in place, we took a series of wavelength calibration
arcs using a HeNeAr lamp before each transit time-series observation. The
narrower slit width of the calibration mask increased the spectral resolution of
the wavelength calibration as well as avoided saturation of the CCD from the arc
lamps. We took a sequence of high SNR flats with a quartz lamp through the
science mask to characterize the pixel-to-pixel variations in the CCD. We ended
up not applying a flat-field correction to the science images after finding in
all previous ACCESS studies (\citealt{rackham2017}, \citealt{espinoza2019};
\citealt{bixel2019}) that flat-fielding introduces additional noise in the data
and does not improve the final results.

\begin{figure*}[htb] \centering
    \includegraphics[width=\linewidth]{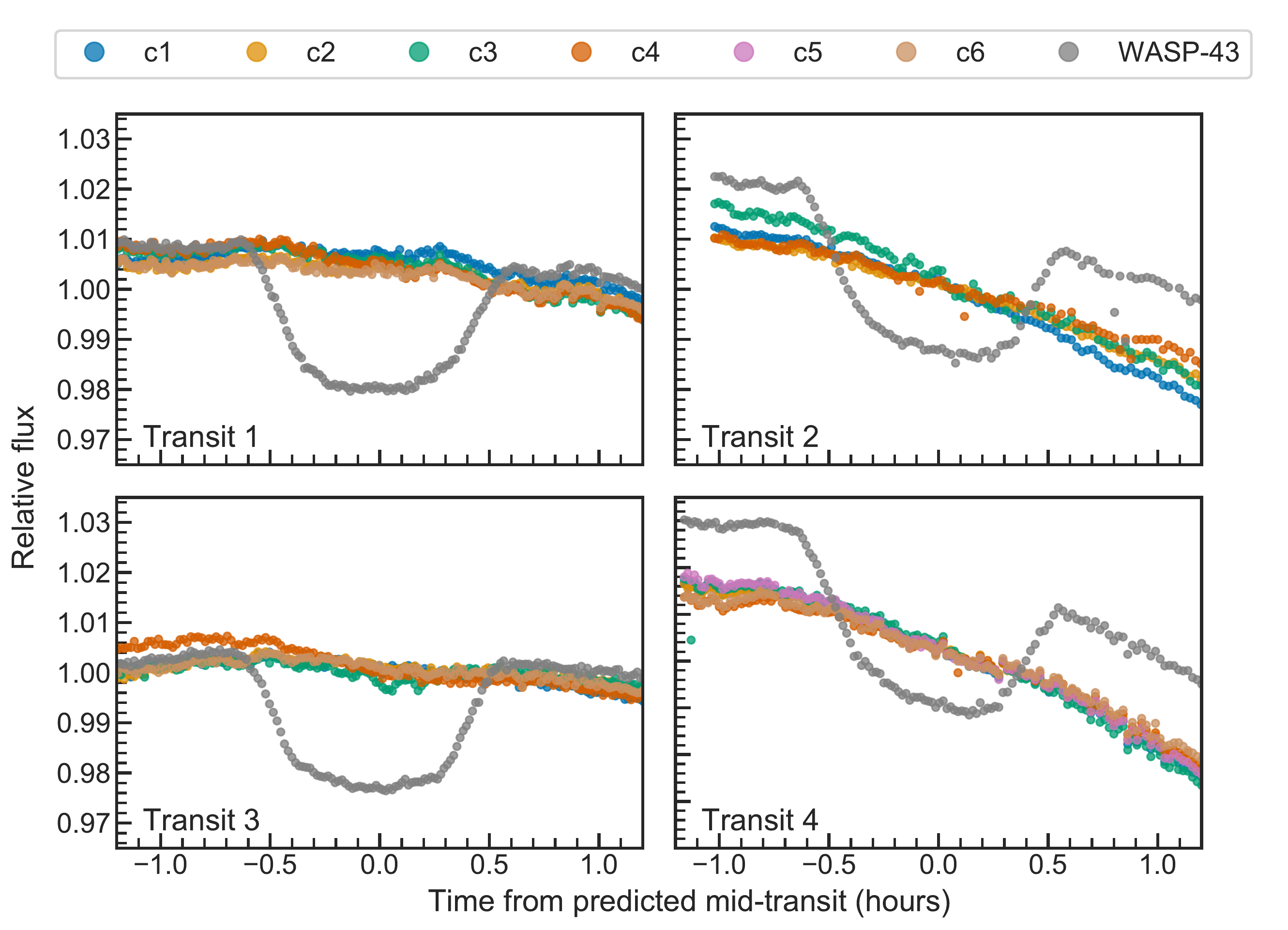}
    \caption{Raw integrated white light curve flux of WASP-43 (grey) and
	    comparison stars (color) observed with IMACS, centered 1 hour around
	    the predicted mid-transit time.  We calculated the predicted
	    mid-transit times with Swarthmore College's online transit finding
	    tool (\url{https://astro.swarthmore.edu/transits.cgi}).}
    \label{fig:raw_flux} 
\end{figure*} 
\begin{figure*}[htb] 
    \centering
    \includegraphics[width=\linewidth]{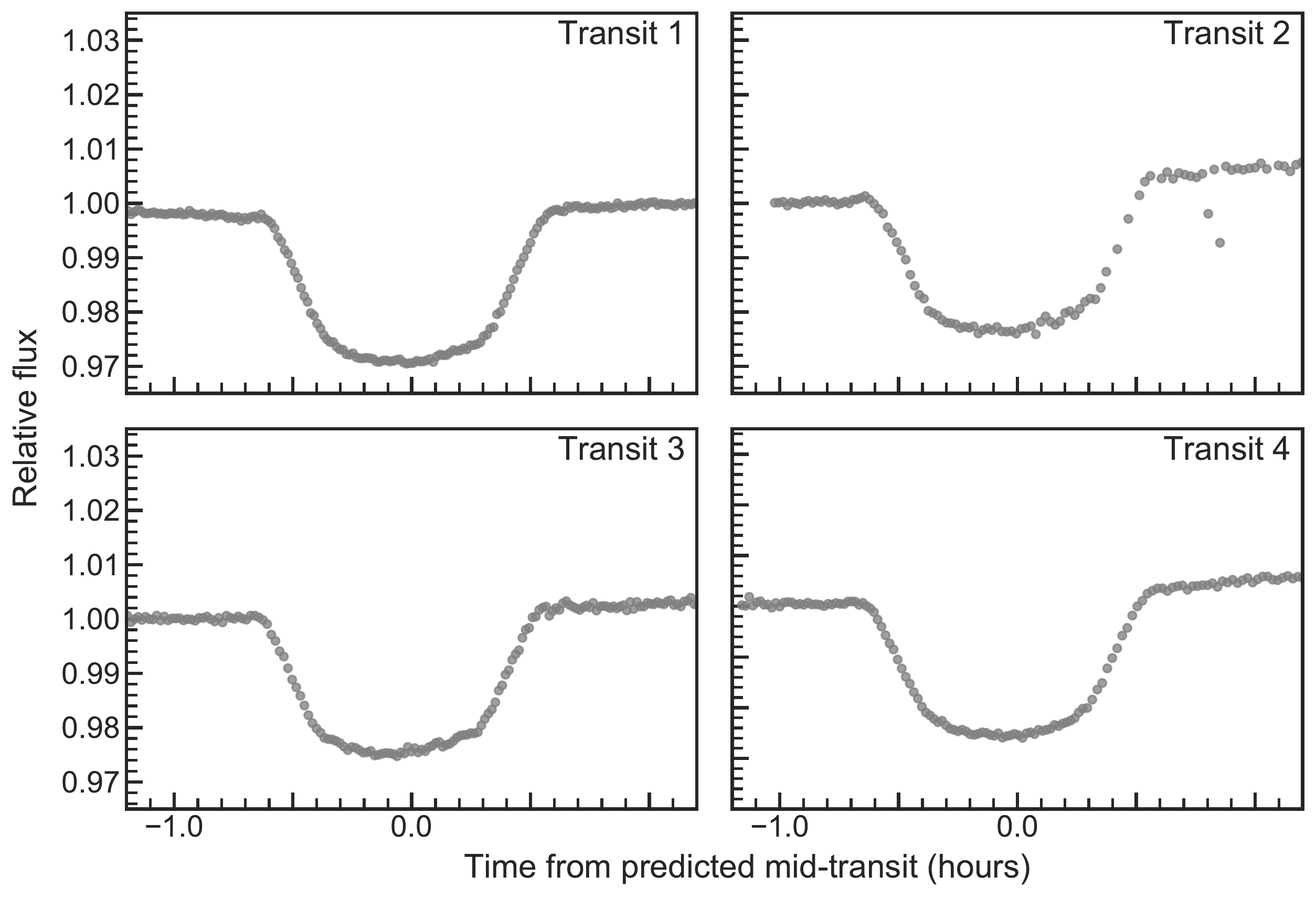}
    \caption{Raw white light curve of WASP-43b divided by the sum of good
	    comparison star flux each night. This informed which comparison 
	    stars and data points to omit from the study.}
    \label{fig:divided_raw_flux}
\end{figure*} 
\begin{figure*}[htb]
    \centering
    \includegraphics[width=\linewidth]{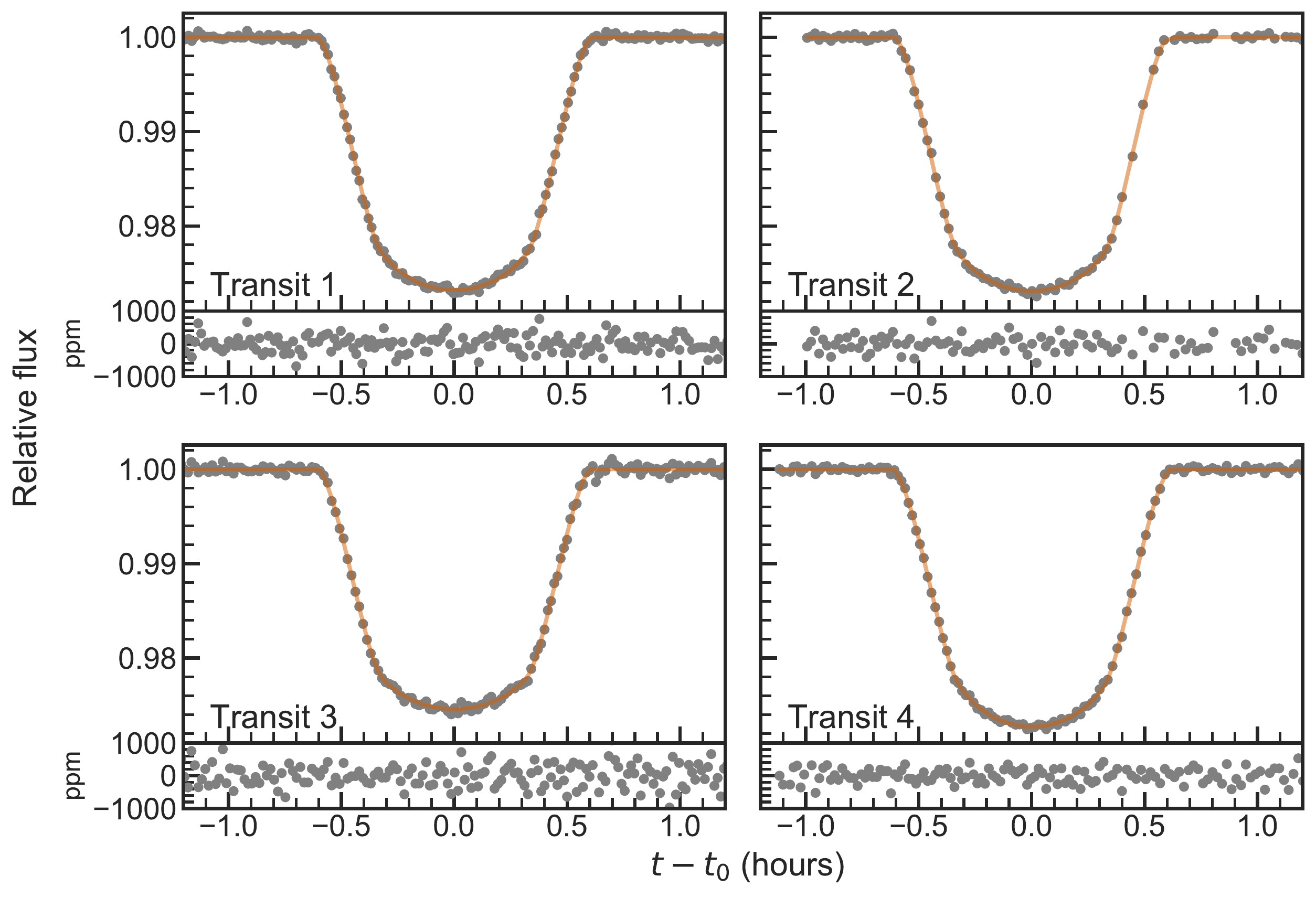}
    \caption{GP+PCA detrended white light curves (grey) and associated models
	    (solid line) for a representative wavelength coverage of
	    \SIrange{5300}{9050}{\angstrom}. We fit for the mid-transit time
	    $t_0$ each transit and used it to perform binned light curve fitting
	    later in the analysis. We center our data 1 hour around $t_0$. 
	    The parameters for each transit fit are given in Table~\ref{tab:wlc_GP}. The associated corner plots are shown in Figures~\ref{fig:corner_wlc_ut150224_GP}~--
	    \ref{fig:corner_wlc_ut180603_GP} of the Appendix}.
    \label{fig:wlc_GP} 
\end{figure*}

\section{Data reduction and light curve analysis} \label{sec:data_red_lc_analysis} 
\subsection{Reduction pipeline}
We reduced the raw data using the ACCESS pipeline described previously
(\citealt{rackham2017}, \citealt{espinoza2019}, \citealt{bixel2019}). The
detailed functions of the pipeline, including standard bias and flat
calibration, bad pixel and cosmic ray correction, sky subtraction, spectrum
extraction, and wavelength calibration are described in detail in
\cite{jordan2013} and \cite{rackham2017}. We briefly summarize the data
reduction here.

We applied the wavelength solution found with the arc lamps to the first science
image and the remaining science image spectra were then cross-correlated with
the first's. We calibrated the spectra from all stars to the same reference
frame by identifying shifts between \halpha\ absorption line minimum of the
median spectra and air wavelength of \halpha, and interpolated the spectra onto
a common wavelength grid using b-splines. We aligned all spectra to within
\SI{2}{\angstrom}, which is less than the average resolution element of
\SI{4.7}{\angstrom}, assuming an average resolution of $R=1,200$ and shortest
wavelength coverage of \SI{5600}{\angstrom} on IMACS. We also subtract scattered
light within IMACS along each slit for every exposure, as described by
\citet{espinoza2019}. 

The final results are sets of wavelength calibrated and extracted spectra for
the target and each comparison star that can be used to produce integrated
(white light, Figure~\ref{fig:raw_flux}) or spectroscopically binned light
curves. The series of white light curves produced in this fashion informed which
comparison stars to omit in the rest of the analysis on a per dataset basis.
Based on the deviations of each comparison star's flux from the general trend of
WASP-43b's flux in Figure \ref{fig:raw_flux}, we omitted comparison star 5 from
the first 2015 dataset (ut150224, Transit 1), comparison stars 5 and 6 from the
second 2015 dataset (ut150309, Transit 2), comparison star 5 from the 2017
dataset (ut170410, Transit 3), and no comparison stars from the 2018 dataset
(ut180603, Transit 4). 

With this set of good comparison stars established, we made a normalized white
light curve $(\Fdiv)$ for each observation by dividing the target WASP-43b flux
$\Ftarg$ by the sum of the $N$ good comparison star fluxes $(\Fcomp)$:
\begin{align} 
    \Fdiv = \frac{\Ftarg}{\sum_n^N\Fcomp}
\end{align} 
to produce the curves shown in Figure~\ref{fig:divided_raw_flux} for
each dataset. We omitted the remaining outliers (points with fluxes
deviating at least $2\sigma$ from the median flux of 10 neighboring points)
from the rest of our detrending procedures. Because the different wavelength bin
schemes we explored (see section~\ref{sec:binschemes}) sometimes had
different outlier points in their respective divided white light curves, we
omitted a super-set of outlier points from all datasets to maintain uniformity
and reduce systematics. 

We detrended our original data, with the selected data points and comparison
stars for each transit epoch selected, using three different methods described
in Section~\ref{sec:lc_analysis}. 

\subsection{Light curve analysis} \label{sec:lc_analysis} 
We applied three detrending and transit-fitting methods to our white light and
wavelength binned data to test our results and verify that they were not
dependent on the method used. The three detrending methods we used were:
polynomial wavelet (poly), polynomial wavelet followed by a common mode
correction (poly+CMC), and Gaussian process combined with principal component
analysis (GP+PCA). 

\subsubsection{Polynomial wavelet detrending} \label{sec:poly} 
We performed a simultaneous transit model and systematics detrending fit on each
of the transit white light curves shown in Figure~\ref{fig:divided_raw_flux}
using our Markov Chain Monte Carlo (MCMC) code, described in detail in
\cite{rackham2017}. To briefly summarize, the divided light curve model 
$F(t, \bt)$ can be written as:
\begin{align}
     F(t, \bt, \ba) = f(\bt)P(t, \ba),
\end{align} 
where $t$ is time, $f(\bt)$ is the analytic transit model described in
\citet{mandel_agol2002}, $\bt$ is the vector of orbital and transit parameters
($e$, $a/\rs$, $i$, $\omega$, $T$, $t_0$, $b$, $u_1$, $u_2$, $\rp/\rs$), defined
in Table~\ref{tab:orb_params}, and $\ba$ is a vector of polynomial coefficients
($\alpha_0$, $\alpha_1$, \dots, $\alpha_m$), where:
\begin{align} 
    P(t, \ba) = \sum_{m=0}^M \alpha_m t^m,
\end{align}
for an $M$th order polynomial that we fit to the out-of transit (OOT) flux of
$\Fdiv$. We assumed a quadratic stellar limb darkening profile and sampled the
limb darkening coefficients $u_1$ and $u_2$ according to \cite{kipping2013} to
allow uninformative (uniform) priors to be placed on the transformed
coefficients while avoiding the risk of sampling non-physical values. We fit the
divided light curve model to determine the most likely values for $\ba$.  We
obtained the final detrended model $(\Fdet)$ by dividing through by the OOT
best-fit flux model $P(t, \ba)$, such that:
\begin{align}
    \Fdet = \frac{\Fdiv}{P(t, \ba)}.
\end{align}

We performed a simultaneous MCMC fitting of transit parameters with
\texttt{PyMC} \citep{pymc}. The likelihood was determined through the wavelet
method described in \cite{carter_winn2009}. The fitted parameters include the
mid-transit time $(t_0)$, planet-to-star radius ratio $(\rp/\rs)$, three
coefficients $(a_i,\ i=0,1,2)$ for the second-order polynomial used to fit the
baseline out-of-transit trend, two parameters for the transformed quadratic
limb-darkening coefficients $(q_1, q_2)$, one noise parameter $(\sigma_\text w)$
for uncorrelated ``white'' noise, and one parameter $(\sigma_\text r)$ for
correlated ``red'' noise. We found that a second order polynomial fit the OOT
flux better than a lower order function.

We sampled $t_0, \sigma_\text w, \sigma_\text r, q_1$, and $q_2$ with uniform
priors, $\alpha_i$ with a Gaussian prior with a width set by bootstrapping the
uncertainty on $\alpha_i$ following \citet{rackham2017}, and $\rp/\rs$ with a
Gaussian prior with spread $5\times\sigma_{\rp/\rs}$, where $\rp/\rs$
and $\mathbf{\sigma_{\rp/\rs}}$ is used from the literature. We used
five chains, each composed of 100,000 steps, and started them at the estimated
location of maximum a posteriori probability (MAP), using an additional 30,000
steps for burn-in. We thinned the chains by sampling them at $10\times$ their
autocorrelation function (ACF) half-life before combining them to produce the
final posterior distributions. Each Bin Scheme has a corresponding source for
the fixed system parameters, and we list them in Table~\ref{tab:orb_params}. We
fit the white light curve while keeping the mid-transit time $t_0$ and average
transit depth $\rp/\rs$ free. 

We produced the binned light curves following the same procedure for producing
the WLCs, with the only difference being that we kept the mid-transit time $t_0$
found in the WLC analysis fixed when performing the fits for $\rp/\rs$ in each
wavelength bin for Bin Schemes 1, 2, and 3, described in
section~\ref{sec:binschemes}. 

\subsubsection{Common-mode correction}
Following \cite{sedaghati2016}, we divided the detrended WLC obtained in the
polynomial detrending method by its best-fit model to produce a common-mode
correction (CMC) residual. We then divided that residual through the binned
light curves to remove wavelength independent variations. Next, we applied the
CMC correction to each Bin Scheme. 

\subsubsection{Gaussian process and principal component analysis} \label{sec:gp_pca}
Gaussian Process (GP) regression is a powerful tool for modeling data in the
machine learning community \citep{rasmussen2005} that has started to gain more
popularity in the exoplanets field (see, e.g., \citealt{gibson2012},
\citealt{aigrain2012}). \cite{gibson2012} provides a good overview to this
methodology applied to exoplanet transit light curves. Applying this methodology
for a collection of $N$ measurements $(\bm f)$, such as the flux of a star
measured over a time series, the log marginal likelihood of the data can be
written as:
\begin{align*}
    \log\mathcal L(\bm r | \textbf{\textsf X},\bm\theta,\bm\phi)
    = &-\frac{1}{2}\bm r^\top\Sigma^{-1}\bm r
    -\frac{1}{2}\log\left|\Sigma\right| \\
      &-\frac{N}{2}\log(2\pi)\quad, \numberthis
\end{align*}
where $\bm r \equiv \bm f - T(\bm t, \bm\phi)$ is the vector of residuals
between the data and analytic transit function $T$; $\textbf{\textsf X}$ is the
$N\times K$ matrix for $K$ additional parameters, where each row is the vector
of measurements $\bm x_n = (x_{n,1}, \cdots x_{n,K})$ at a given time $n$;
$\bm\theta$ are the hyperparameters of the GP; $\bm\phi$ are the transit model
parameters; and $\Sigma$ is the covariance of the joint probability distribution
of the set of observations $\bm f$. In our analysis, we used six systematics
parameters: time, full width at half maximum (FWHM) of the spectra on the CCD,
airmass, position of the pixel trace through each spectra on the chip of the
CCD, sky flux, and shift in wavelength space of the trace. We used the Python
package \texttt{batman} \citep{kreidberg2015} to generate our analytic transit
model. From here, the log posterior distribution $\log\mathcal
P(\bm\theta,\bm\phi|\bm f,\textbf{\textsf X})$ can be determined by placing
explicit priors on the maximum covariance hyperparameters and the scalelength
hyperparameters. From $\mathcal P$, the transit parameters can then be inferred
by optimizing with respect to $\bm\theta$ and $\bm\phi$.

We accomplished the above optimization problem with the Bayesian inference tool,
\texttt{PyMultiNest} \citep{buchner2014}, and computed the log likelihoods from
the GP with the \texttt{george} \citep{george} package.  We implemented this
detrending scheme by simultaneously fitting the data with an exponential squared
kernel for the GP under the assumption that points closer to each other are more
correlated than points farther apart. The PCA methodology follows from
\citet{jordan2013} and \citet{espinoza2019}, where $M$ signals, $S_i(t)$, can be
extracted from $M$ comparison stars and linearly reconstructed according to the
eigenvalues, $\lambda_i$, of each signal. This allows for the optimal extraction
of information from each comparison star to inform how the total flux of WASP-43
varies over the course of the night. We Bayesian model averaged (BMA) the
principal components together, which were determined by fitting with one, then
two, up to $M$ principal components, to create the final detrended WLC and model
parameters of interest. We present the associated WLCs in 
Figure~\ref{fig:wlc_GP}, the best fit parameters in Table~\ref{tab:wlc_GP}, and
associated corner plots in 
Figures~\ref{fig:corner_wlc_ut150224_GP}--\ref{fig:corner_wlc_ut180603_GP} in the
Appendix. Based on the quality of the fits, discussed in Section~\ref{sec:bin_kreidberg},
and to streamline our work, we show only the results of this method for the 
transmission spectra that informed our retrieval analysis.

We applied the same methodology on a wavelength bin by wavelength bin basis to
produce the simultaneously fitted light curves for Bin Scheme 3 and create our
final transmission spectrum, shown in Figure~\ref{fig:tspec_combined}. We used
the open source package, \texttt{ld-exosim}
\footnote{\url{https://github.com/nespinoza/ld-exosim}}, to determine that a
square-root limb darkening law was the most appropriate for WASP-43, and
incorporated this into our GP analysis. We present the final transmission
spectrum using this method in Figure~\ref{fig:tspec_combined}.

\begin{deluxetable*}{lCCC}[htb]
    \caption{Literature system parameters.}
    \label{tab:orb_params} 
    \tablehead{\colhead{Bin Scheme} & \colhead{1} &
    \colhead{2} & \colhead{3}} 
    \startdata 
    eccentricity $(e)$                        & 0                           & 0                            & 0     \\ 
    semi-major axis/stellar radius $(a/\rs)$  & 4.867\pm 0.023              & 4.752\pm 0.066               & 4.872 \\ 
    inclination (radians) $(i)$               & 1.426\pm 0.0056             & 1.433\pm 0.00175             & 1.433 \\ 
    planet/stellar radius uncertainty $(\sigma_{\rp/\rs})$ & 0.0018 & 0.00145 & 0.00043^{(a)}                      \\
    longitude of periastron $(\omega)$        & \pi                & \pi                          & \pi                \\ 
    period (days) (T)                         & 0.813473978\pm\SI{3.5E-8}{} & 0.81347459 \pm \SI{2.1E-7}{} & 0.81347436 \\ 
    Reference                                 & \text{\cite{hoyer2016}}     & \text{\cite{murgas2014}}     & \text{\cite{kreidberg2014}} 
    \enddata 
    \tablecomments{Literature system parameters corresponding to each Bin Scheme. 
    (a) We used $\sigma_{\rp/\rs}$ from \citet{hoyer2016} combined
    transit data because they are based on ground-based values while values from
    \citet{kreidberg2014} are space based.} 
\end{deluxetable*}

\begin{deluxetable*}{CcRRRR}[htb]
    \caption{Fitted WLC values from GP+PCA detrending method shown in
            Figure~\ref{fig:wlc_GP}. We share the associated corner plots in
            Figures~\ref{fig:corner_wlc_ut150224_GP}~--\ref{fig:corner_wlc_ut180603_GP}
            of the Appendix. Note: we computed transit depths directly from $\rprs$.}
    \label{tab:wlc_GP}
    \tablehead{\colhead{parameter} & \colhead{definition} & \colhead{Transit 1} & \colhead{Transit 2} &
    \colhead{Transit 3} & \colhead{Transit 4}}
    \startdata 
    \rp/\rs          &      planet radius / star radius           & 0.15854^{+0.00079}_{-0.00074}    & 0.15800^{+0.00218}_{-0.00278}    & 0.15436^{+0.00167}_{-0.00164}    & 0.16030^{+0.00127}_{-0.00118}    \\
    d &	transit depth (ppm) & 25134 \pm 250 & 24963 \pm 880 & 
    23828 \pm 516 & 25695 \pm 406 \\
    t_0-2450000      & mid-transit (JD) & 7077.72325^{+0.00004}_{-0.00004} & 7090.73888^{+0.00008}_{-0.00008} & 7854.59100^{+0.00008}_{-0.00008} & 8273.53019^{+0.00005}_{-0.00005} \\
    P                & period (days)    & 0.81347^{+0.00000}_{-0.00000}    & 0.81347^{+0.00000}_{-0.00000}    & 0.81347^{+0.00000}_{-0.00000}    & 0.81347^{+0.00000}_{-0.00000}    \\
    a/\rs            &       semi-major axis / star radius           & 4.92738^{+0.02856}_{-0.02850}    & 4.97737^{+0.05644}_{-0.06297}    & 4.90935^{+0.05881}_{-0.05815}    & 4.85836^{+0.03458}_{-0.03367}    \\
    b                & impact parameter & 0.65644^{+0.00698}_{-0.00720}    & 0.65270^{+0.01433}_{-0.01372}    & 0.67153^{+0.01335}_{-0.01576}    & 0.66218^{+0.00822}_{-0.00818}    \\
    i                & inclination      & 82.34509^{+0.11731}_{-0.11874}   & 82.46517^{+0.23358}_{-0.25848}   & 82.13717^{+0.26874}_{-0.24363}   & 82.16642^{+0.14519}_{-0.14579}   \\
    q_1              & LD coeff 1       & 0.70515^{+0.19778}_{-0.23016}    & 0.60216^{+0.23895}_{-0.17155}    & 0.50975^{+0.30047}_{-0.19881}    & 0.72406^{+0.18106}_{-0.19668}    \\
    q_2              & LD coeff 2       & 0.38994^{+0.10395}_{-0.19375}    & 0.30680^{+0.17608}_{-0.18818}    & 0.41212^{+0.21415}_{-0.24787}    & 0.31602^{+0.10556}_{-0.16639}    
    \enddata 
\end{deluxetable*}

\subsubsection{White light curve and Binning Schemes}\label{sec:binschemes}
We applied the detrending methods described in
Sections~\ref{sec:poly}--~\ref{sec:gp_pca} to the following wavelength
binning schemes in our analysis:  
\begin{itemize}
    \item Bin Scheme 1: A set of uniform bins centered around the air
        wavelength values of key spectral features (\ion{Na}{1}-D, \halpha,
        \ion{K}{1}, \ion{Na}{1}-8,200) to produce a transmission spectrum
        focused around these features; 
    \item Bin Scheme 2: identical binning and system parameters to
        \cite{murgas2014} to directly compare our transmission spectra with the
        ones presented in that study;
    \item Bin Scheme 3: similar \SI{230}{\angstrom} binning and system
        parameters to \cite{kreidberg2014} to combine our visual measurement
        with their NIR measurements made with HST. We used this binning scheme
        to perform atmospheric retrievals described in
        Section~\ref{sec:bin_kreidberg}. 
        We centered \SI{230}{\angstrom} bins around the vacuum wavelength locations 
        of \ion{Na}{1}-D, \halpha,\ \ion{K}{1}, and \ion{Na}{1}-8,200 using smaller bins when necessary 
        to have at least two wavelength bins between each feature.
\end{itemize}
We applied the poly+CMC detrending method described in
Section~\ref{sec:poly} to all three Bin schemes and applied the GP+PCA described
in Section~\ref{sec:gp_pca} to Bin Scheme 2 (\SI{25}{nm}) and Bin Scheme 3
because of their similar wavelength binning and wavelength coverage in the
visual (\SI{25}{nm} vs. \SI{23}{nm}). This allowed us to directly compare
detrending methods between our study and \cite{murgas2014}.

We discuss each binning scheme in more detail in
Sections~\ref{sec:bin_species},~\ref{sec:bin_murgas}, and
\ref{sec:bin_kreidberg}. We also include the literature values for system
parameters used in each Bin Scheme in Table~\ref{tab:orb_params}. For
Bin Scheme 1, where no associated literature values are being used for
comparison, we adopt the most up to date values from \cite{hoyer2016}.

\section{Stellar Activity} \label{sec:stell_act}
Before combining the transmission spectra from each night, we first considered
the impact of stellar photospheric heterogeneity, which can have an observable
effect on transmission spectra \citep{pont2008, pont2013, sing2011, oshagh2014,
zhang2018}, even if magnetically active regions are not occulted by the
transiting exoplanet \citep{mccullough2014, rackham2018, rackham2019, apai2018}.
Qualitatively, global variations in stellar activity could manifest themselves
as an overall dimming or brightening of the star, which could lead to
significant variations in transit depths. Changes in photometric activity roughly 
correlate with the covering fraction of starspots, which in turn can modulate the 
luminosity of the star and impact observed transit depths \citep{berta2011}. 
Furthermore, those variations can be
wavelength dependent, leading to slopes with spurious spectral features in the
transmission spectrum.

The white light transit depths we observed (Table~\ref{tab:wlc_GP})
varied by as much as \SI{1869}{ppm} between transits. To account for this
offset between the datasets, we investigated the contribution due to stellar
activity.  Changes in photometric activity roughly correlate with the covering
fraction of starspots, which in turn can modulate the luminosity of the star and
impact observed transit depths \citep{berta2011}. To assess brightness variation
of WASP-43 over the time frame of our observations, we used
15~Feb~2012~--~21~May~2018 activity data from 990 out-of-transit V-band images
of WASP-43b taken by Ohio State University's All-Sky Automated Survey for
Supernovae\footnote{\url{https://asas-sn.osu.edu}} (ASAS-SN) program
\citep{shappee2014,kochanek2017}.

\begin{figure}[htb]
    \centering
    \includegraphics[width=\linewidth]{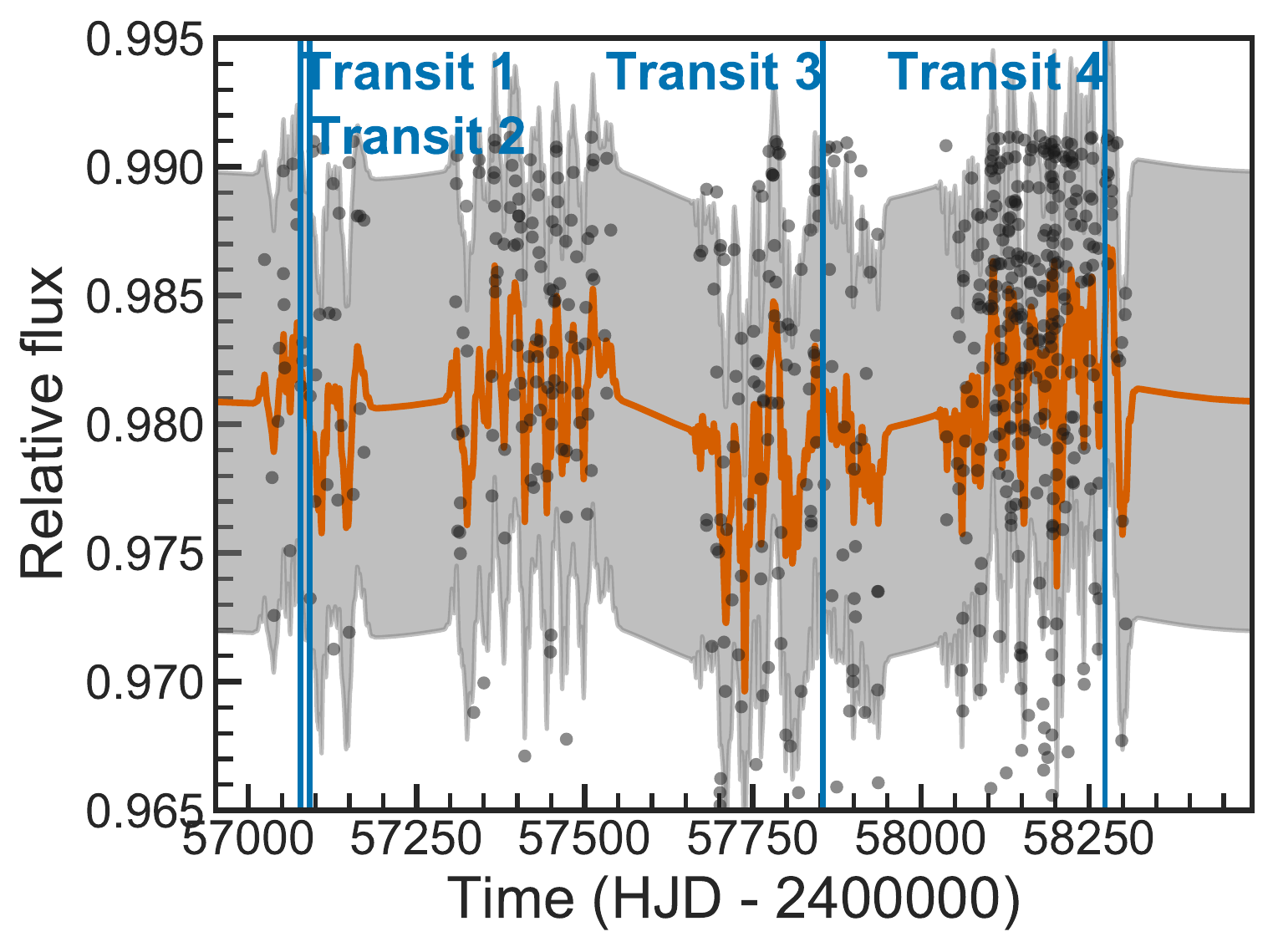}
    \caption{Ground-based photometric observations of WASP-43 from ASAS-SN (grey
	    points) during Transit 1, Transit 2, and Transit 3 transit epochs
	    (blue vertical lines). The data are flux relative to the average
	    brightness of comparison stars. The Gaussian process regression
	    model (red) and $1\sigma$ uncertainty (gray region) fit to ASAS-SN
	    data are also overplotted.}
\label{fig:phot_mon_full}
\end{figure}

The ASAS-SN photometric activity was sampled much more coarsely than our transit
observations, so we used a regression routine \citep{alam2018} to fit the data
and estimate the amplitude of the photometric variation induced by stellar
activity during each of the four transit epochs of WASP-43b. Following
\cite{alam2018}, we used a negative log likelihood kernel $(\ln\textbf L)$ for
the objective function given by:
\begin{align*}
\ln(\mathbf L) =
&-\frac{n}{2}\ln(2\pi) - \frac{1}{2}\ln(\det\textbf K) \\&- \frac{1}{2}(\textbf
y - \mu)^\top\textbf{K}^{-1}(\textbf y - \mu)\numberthis\quad,
\end{align*}
where $\textbf y$ is the data, $\mu$ is the model, $n$ is the number of
observations, and \textbf K is the covariance matrix. \textbf K describes the
correlation weight between all possible pairs of photometric measurements and is
populated with the GP kernel to quantify the correlation of pairs of
observations. We used a gradient based optimization routine to find the best fit
hyperparameters and used the 15.6 day stellar rotation period from
\cite{hellier2011}. Figure~\ref{fig:phot_mon_full} shows the GP regression model
for the relevant ASAS-SN data.

\begin{figure*}[htb]
    \centering
    \includegraphics[width=\linewidth]{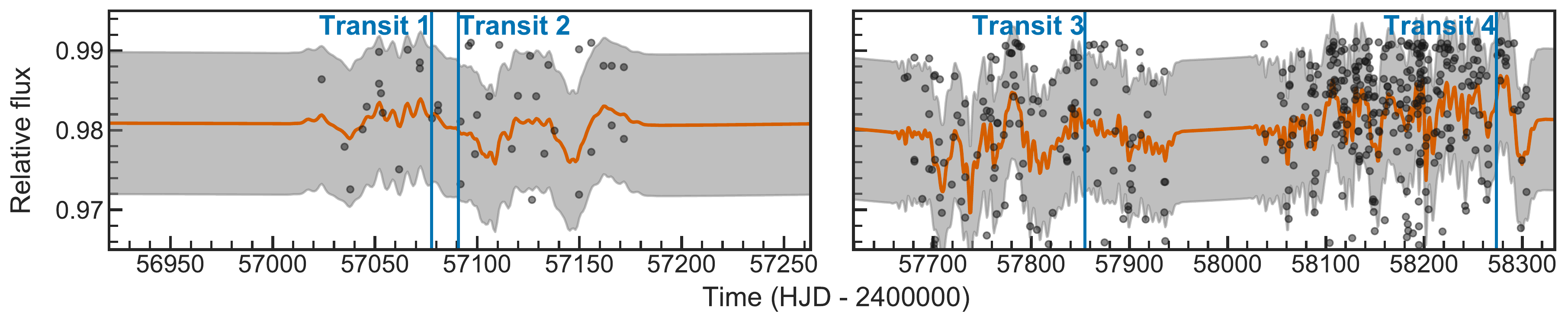}
    \caption{ASAS-SN photometric data from Figure~\ref{fig:phot_mon_full}} centered around all four transit epochs.
    Because the data is too sparse to cover even a single transit epoch, we
    do not use it to quantify the contribution of stellar activity to the
    transmission spectrum, but instead rely on the formalism discussed in
    Section~\ref{sec:bin_kreidberg}.
    \label{fig:phot_mon_zoomed}
\end{figure*}

Figure~\ref{fig:phot_mon_full} shows the complete ASAS-SN light curve, and
Figure~\ref{fig:phot_mon_zoomed} details the ASAS-SN photometry near each of our
transit epochs. Overall, the relative flux from the photometric monitoring
varies by as much as 3\% from the median value obtained from the GP.
Observations with ASAS-SN are too coarse to effectively sample the photometric
activity during times of transits. However, this data still gives us a rough
idea of differences in stellar flux between epochs.

We conclude that the photometric activity data alone are not enough to constrain
the contribution of unocculted heterogeneities on the surface of WASP-43b to the
resulting transmission spectrum. Nonetheless, we argue that changes in disk
coverage by unocculted heterogenieties likely drive the white-light light curve
depth variations that we observe between transit epochs. For this reason, we
calculate and apply transit depth offset corrections as described in
Section~\ref{sec:comb_nights} before building the final transmission spectrum.
In Section~\ref{sec:bin_kreidberg} we model the possible contribution of an
unocculted heterogeneous photosphere to the resulting transmission spectrum
without relying on photometric monitoring data.

\section{Transmission spectrum}\label{sec:tspec}
\subsection{Combining nights}\label{sec:comb_nights}
The uncertainties in transit depths from our individual transits range from 
\SIrange{250}{880}{ppm}, which is not enough to detect the atmosphere of WASP-43b 
(see Table~\ref{tab:wlc_GP}). For example, a hydrogen dominated composition 
for the atmosphere of WASP-43b would produce a signal $\Delta D$ of 435 ppm at 5 
scale heights \citep[based on eq.~11 of][and using the planet and star parameters
given in Section~\ref{sec:intro}]{Miller_Ricci_2009}. Therefore, we needed to
combine the transmission spectra from the four transits to be sensitive to
atmospheric features of the planet.

Transit depths between different epochs varied as much as 1867 ppm
(see difference in depths between Transits 3 and 4 in
Table~\ref{tab:wlc_GP}), an effect that we attribute to stellar variability.
We also expect some variability in observed transit depths for two reasons: 
i) stellar activity is stronger in the visual portion of the spectrum
relative to in the IR, and ii) because of the large period of time over
which we collected our data. For example, \cite{kreidberg2014} observed
six transits over 1 month, while our data spans four transits over
3 years. This leaves ample time for the star's intrinsic brightness to
change due to surface stellar heterogeneities and impact measured
transit depths.

To be able to combine the data from each transit epoch, we needed to first
consider potential effects introduced by stellar activity of WASP-43, as discussed 
in Section~\ref{sec:stell_act}. 
Typically, studies have used photometric activity as a proxy for
the presence of these stellar heterogeneities, but this (compounded with the
fact that the photometric activity data we have is not well sampled enough in
time to cover the given transits) has been shown to be insufficient to correct
for these stellar contributions to transit depth variations
\citep{mccullough2014}. Based on the lack of constraints on the contribution of
occulted heterogeneities on the stellar photosphere to the transmission spectrum
discussed in Section~\ref{sec:stell_act}, we averaged the transmission spectrum
from each night together, weighted by the wavelength dependent uncertainty
estimated from the wavelength binned fitting. Before taking this weighted
average we first addressed the apparent offset visible in the resulting
transmission spectra (colored points in Figure~\ref{fig:tspec_combined}). We did
this by subtracting the mean white light transit depth of the four nights from
the transmission spectrum of each night. After applying the offset, we combined
the four transit epochs by averaging the transmission spectra from each night
together, weighted by the uncertainties in the wavelength dependent depths
determined by the fitting. We took the maximum of this asymmetrical uncertainty
to be conservative in our weighting. We applied this methodology to each Bin
Scheme identified in the following section. Effectively, retrievals on the 
resulting transmission spectrum found probe for the average contribution from the 
stellar photosphere over all transit events.

\subsection{Bin Scheme 1: Species dependent binning}\label{sec:bin_species}
\begin{figure*}[htb]
    \centering
    \includegraphics[width=\linewidth]{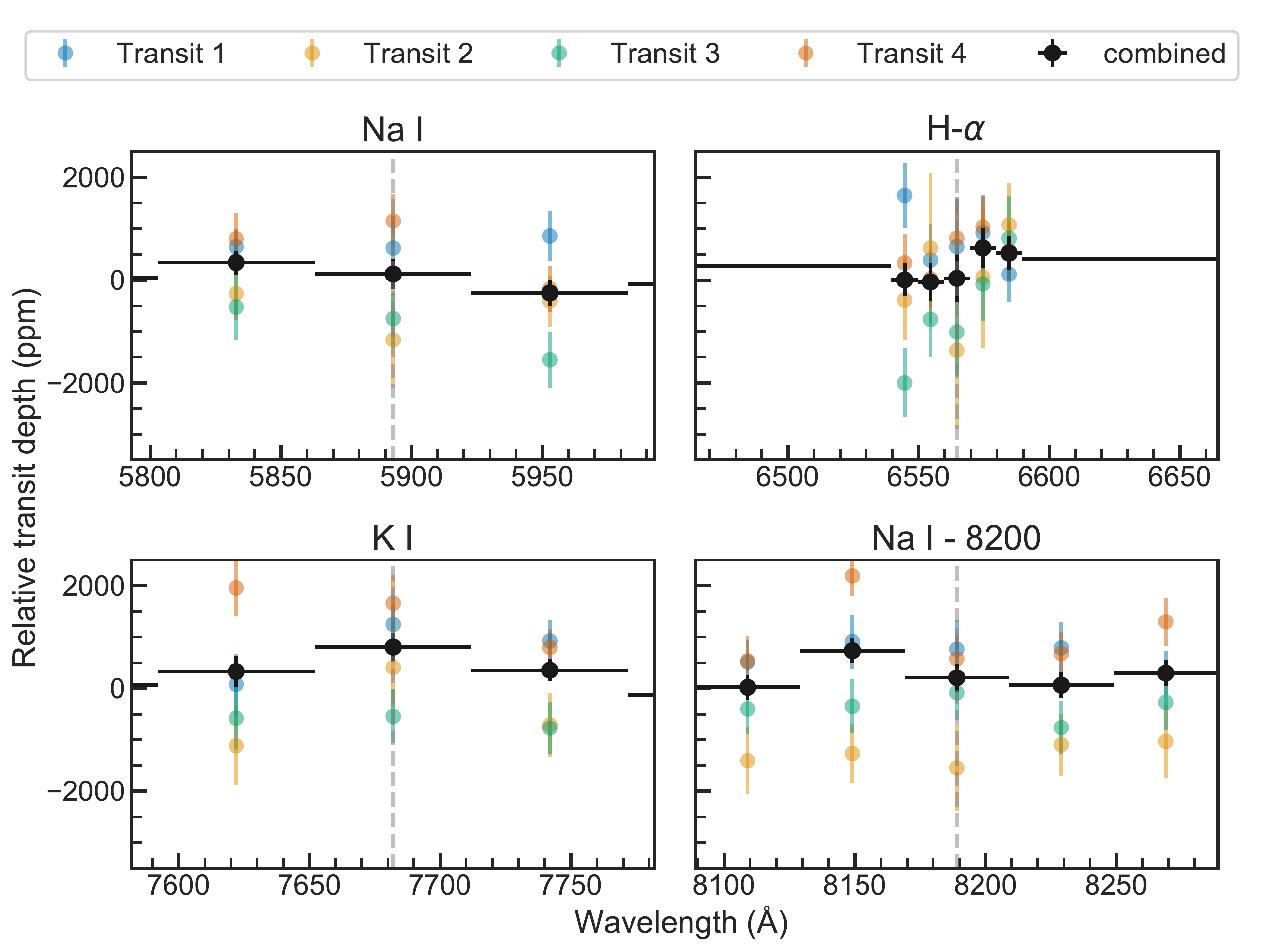}
    \caption{Portions of the transmission spectrum obtained using Bin Scheme 1, 
    centered around \ion{Na}{1}, \ion{K}{1}, and \halpha\ features. 
    Individual nights are shown as colored points, the combined spectrum is shown in 
    black. We made the transmission spectrum sensitive to any potential features that 
    may exist by decreasing the bin size as it approaches the air wavelength of
    each potential species we searched for. We do not detect an excess in transit 
    depth for any of the species.}
    \label{fig:combined_species} 
\end{figure*}

In this binning scheme, we set the wavelength bin sizes based on the absorption
band widths of features of interest, in particular: \ion{Na}{1}-D, \halpha,
\ion{K}{1}, and \ion{Na}{1}-8,200. We set the minimum bin size for a given
feature to be equal to the full width of its observed stellar absorption line,
including the contributions from the wings of the line. This gives a set of 4
bin widths equal to 60, 10, 60, and \SI{40}{\angstrom} for the respective
species listed above. For each species, we mapped a region covering 5 times its
bin width above, centered on the air wavelength to resolve any potential peaks,
and we used larger bins to cover the rest of the spectrum. We produce the
combined spectra in Figure~\ref{fig:combined_species} following this procedure
for the (poly+CMC) detrending scheme.

We observed an apparent peak near the \ion{Na}{1}-8,200 line, but the fact that
the only point far from the baseline is also far from the air wavelength for
this species, indicate that this peak is most likely due to residuals from water
tellurics. We also observed a potential \ion{K}{1} peak 1$\sigma$ above the
median in the bin two \SI{10}{\angstrom} immediately redward of its air
wavelength location, but believe that this is due to residual telluric
absorption as well. Furthermore, we do not detect this peak at all in the GP+PCA
detrended data. We also do not detect an absorption peak near \ion{Na}{1}-D or
\halpha. To set this Bin Scheme apart from the other two Bin Schemes, we used
more up to date system parameters from \citet{hoyer2016}.

\subsection{Bin Scheme 2: Comparison to ground-based study} \label{sec:bin_murgas}
\begin{figure*}[htb]
    \centering
    \includegraphics[width=\linewidth]{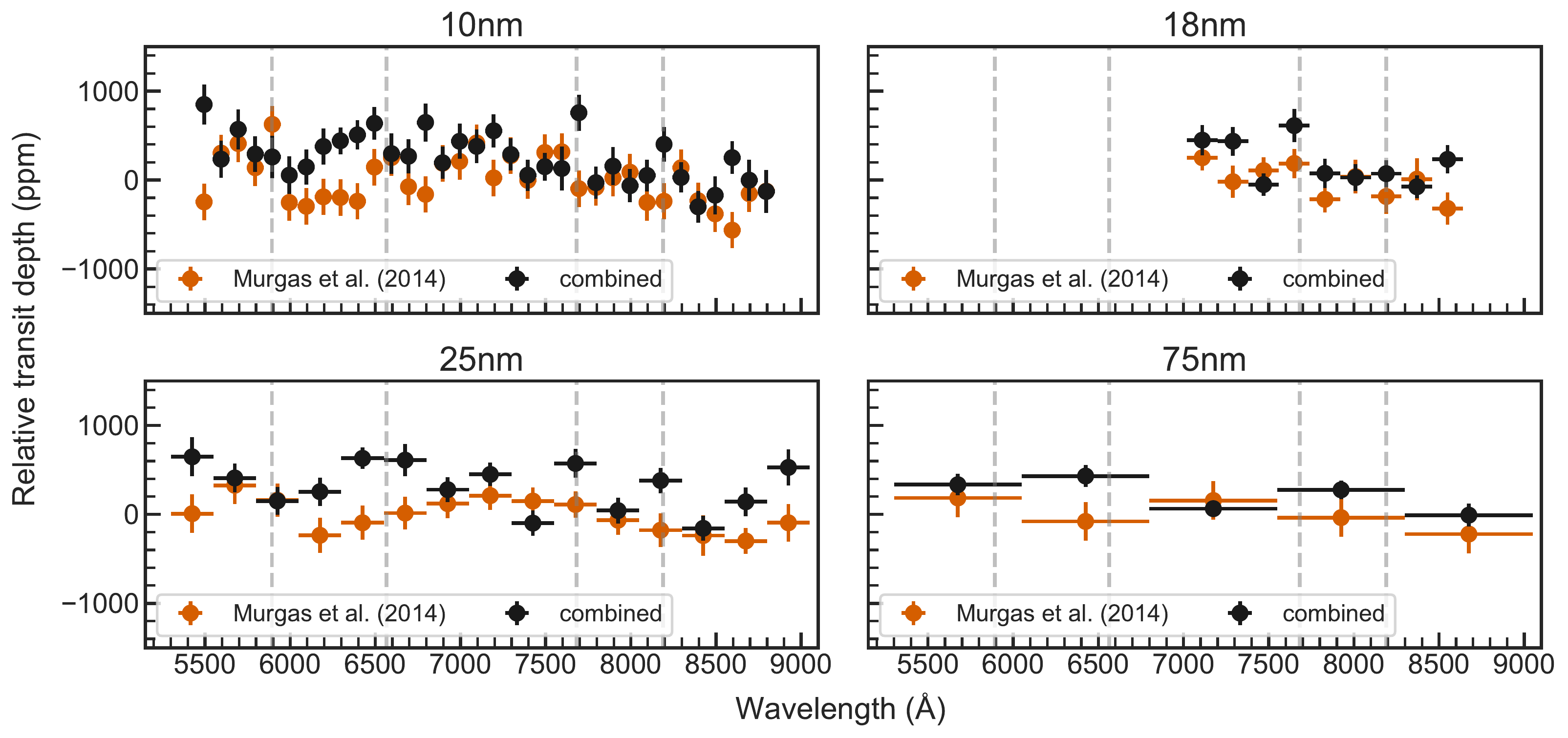}
    \caption{Combined poly+CMC detrended transmission spectrum comparing
	    the four \cite{murgas2014} binning schemes against our own
	    spectrum (black). We were unable to find an associated data
	    table for the \SI{10}{nm} and \SI{75}{nm} bins so
	    we used a manual online digitizer instead
	    (\url{https://apps.automeris.io/wpd/}). There appears to be
	    a peak near \ion{K}{1}, but we consider this to be a
	    spurious detection.  For example, we do not observe this
	    peak in the GP+PCA detrended data. From left to right, the
	    vertical dashed lines mark the air wavelength locations of 
	    \ion{Na}{1}, \halpha, \ion{K}{1}, and \ion{K}{1}-8,200, 
	    respectively.}
    \label{fig:tspec_combined_murgas} 
\end{figure*}

\begin{figure*}[htb]
    \centering
    \includegraphics[width=\linewidth]{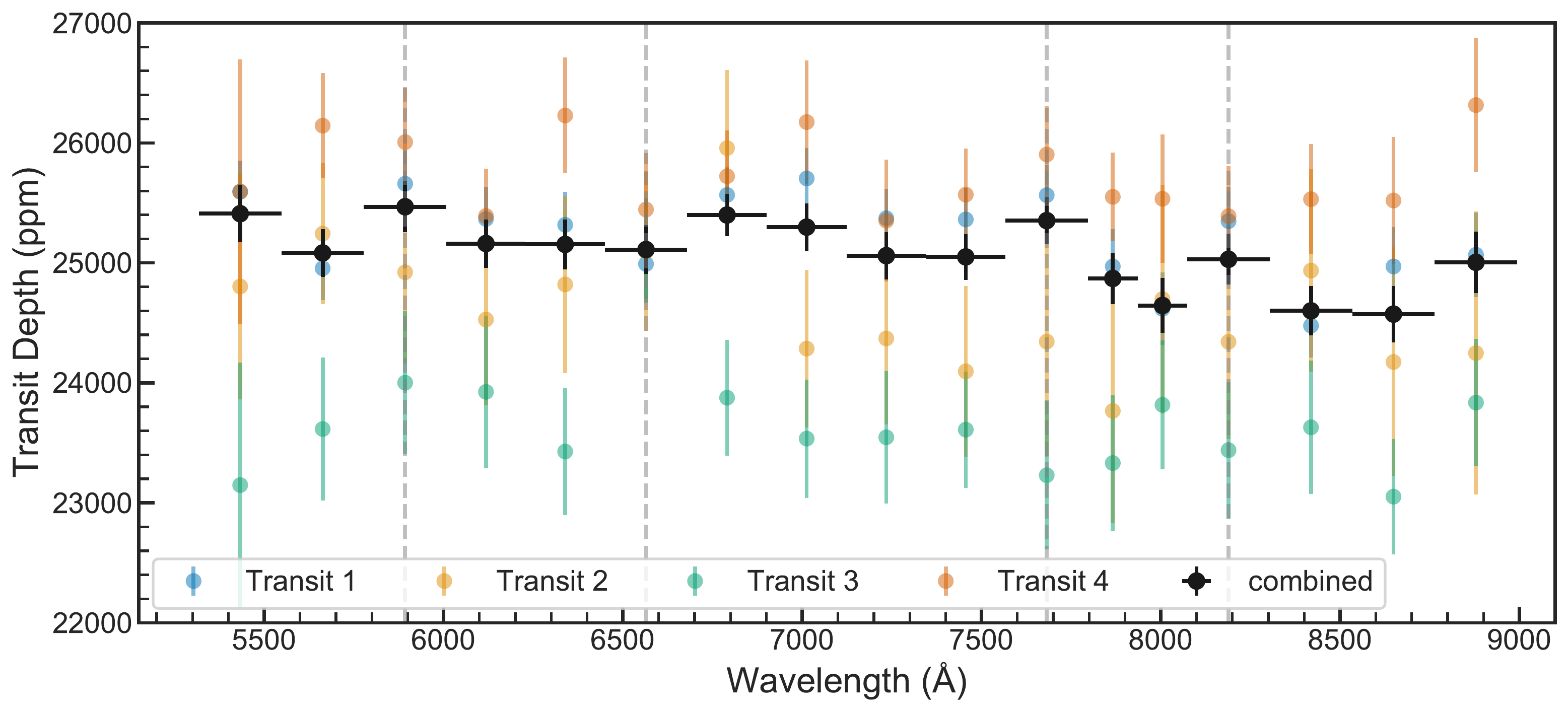}
    \caption{GP+PCA detrended transmission spectrum of Bin Scheme 3. The
	    transit depths are all relative to the weighted mean of the
	    white light depths. From left to right, the vertical dashed
	    lines mark the air wavelength locations of \ion{Na}{1},
	    \halpha, \ion{K}{1}, and \ion{K}{1}-8,200, respectively. We share the table 
	    for the above data in Table~\ref{tab:binned_table_gp} of the Appendix.}
    \label{fig:tspec_combined} 
\end{figure*}

We applied the same detrending and combining methods using the system parameters
reported by \cite{murgas2014} to compare our transmission spectra with those
from their similar study of WASP-43b. We adopted the following four wavelength
bin schemes identified in their work: (i) \SI{100}{\angstrom} bins ranging from
\SIrange{5445}{8845}{\angstrom}, (ii) \SI{250}{\angstrom} bins ranging from
\SIrange{5300}{9050}{\angstrom}, (iii) \SI{750}{\angstrom} bins ranging from
\SIrange{5300}{9050}{\angstrom}, (iv) \SI{180}{\angstrom} bins centered near the
\ion{K}{1} \SI{7665}{\angstrom} and \SI{7699}{\angstrom} doublet, to produce
Figure~\ref{fig:tspec_combined_murgas}.

Qualitatively, the shapes of the combined spectra in each binning scheme tend to
follow the same slight upward curving slope near \SI{7000}{\angstrom} seen in
\cite{murgas2014}. Unlike their finding, we do not observe an excess near
5,892.9 \angstrom\ that would indicate the presence of \ion{Na}{1} in the
atmosphere of WASP-43b.

\subsection{Bin Scheme 3: Combining with NIR study} \label{sec:bin_kreidberg}
We used the same bin width and system parameters (Table~\ref{tab:orb_params})
from \cite{kreidberg2014} to combine their NIR transmission spectrum with our
visual spectrum. We found that although the polynomial+CMC detrending method
tended to produce smaller error bars on average in the transmission spectrum
($\SI{180}{ppm}$ vs. $\SI{210}{ppm}$), GP+PCA does a better job overall at
fitting systematics in the light curves, resulting in an average standard
deviation across the transit epochs of \SI{396}{ppm} vs. \SI{274}{ppm} in the WLC
model residuals, respectively. The average GP+PCA WLC residuals were also
larger than the average white noise of \SI{146}{ppm} estimated in the poly+CMC
detrended WLCs, so the GP+PCA procedure was not over-fitting the data.
For these reasons, we adopted the GP+PCA detrended transmission spectrum,
shown as black circles in Figure~\ref{fig:tspec_combined}, as
our final spectrum (reported in Table~\ref{tab:binned_table_gp}), which we combined with the NIR transmission spectrum from
\cite{kreidberg2014} for the atmospheric retrieval analysis.

\section{Atmospheric Retrieval Analysis} \label{sec:retrieval}
With the final combined visual+NIR transmission spectrum from
Section~\ref{sec:bin_kreidberg}, we searched for signals in the atmosphere of
WASP-43b. We used a Bayesian atmospheric retrieval code based on the same nested
sampling Bayesian inference software we used for performing the GP+PCA
detrending, \texttt{PyMultiNest}. The details of the retrieval code are given in
our previous study of WASP-19b \citep{espinoza2019}, and we only briefly
summarize the methodology here.

Following the semi-analytical formalism from \citet{betremiex2017} and
\citet{heng2017}, we assume an isothermal and isobaric atmosphere, with an
optically thick base region with radius $(\rp/\rs)_0$ and reference pressure
$P_0$, which we interpret as the cloud-top pressure. Above this region is an
optically thin planetary atmosphere with average temperature $T$ that can have
either (i) a set of atomic and molecular species and/or (ii) a scattering haze
defined by $\sigma_\text{haze}(\lambda) =
a\sigma_0(\lambda/\lambda_0)^{\gamma_\text{haze}}$
\citep{macdonald_and_madhusudhan2017}, where $\sigma_0=\SI{5.31E-27}{cm^2}$ is
the Rayleigh scattering cross-section of H$_2$ at the reference wavelength
$\lambda_0=\SI{350}{nm}$, and $a$ and $\gamma_\text{haze}$ are free parameters.
We constrain $\gamma_\text{haze}$ to be between 0 (uniform opacity) and -4
(Rayleigh scattering) to allow for a better constraint on $a$.
Transmission spectra from separate studies can be combined
by retrieving for an offset between the different datasets. A detailed overview
of the retrieval framework is given in Appendix D of \citet{espinoza2019}.

Additionally, we explored the impact of a heterogeneous stellar photosphere on
the observed transmission \citep{pinhas2018} by following the formalism
described by \citet{rackham2018, rackham2019}. To summarise directly from the
schematic in Figure 1 of \citet{rackham2018}: during a transit, exoplanet
atmospheres are illuminated by the portion of a stellar photosphere immediately
behind the exoplanet (the transit chord).  Changes in transit depth must be
measured relative to the spectrum of this light source. However, the lightsource
is generally assumed to be the disk-integrated spectrum of the star. Any
differences between the assumed and actual light sources will lead to apparent
variations in transit depth.

In this framework, stellar contamination of the transmission spectrum from
unocculted star spots and faculae are considered by placing constraints on the
allowed spot and faculae covering fractions using a set of rotating photosphere
models and then translating the covering fractions of potential stellar
contamination in the transmission spectrum. We incorporate this transit light
source (TLS) effect into our retrieval framework with a three-parameter model
for the stellar photosphere to fit this simultaneously with the planet's
atmosphere. The three parameters are: $\Tchord$, the effective temperature of
the transit chord; $\Thet$, the mean effective temperature of the heterogeneous
features not occulted by the transit chord; and $\fhet$, the fraction of the
projected stellar disk covered by these heterogeneous features. The impact of
these heterogeneities on the transmission spectrum are expressed by the
wavelength dependent corrective factor $\epsilon_\lambda$ on the transit depth,
where:
\begin{align*}
\epsilon_\lambda
\equiv
\left[1 - \fhet\left(1-\frac{\Shet}{\Schord} \right)\right]^{-1}
= \left(\frac{\rplam}{\rplamnot} \right)^2 \quad .
\numberthis
\end{align*}
Here $\rplam$ and $\rplamnot$ are the apparent and actual planetary radius
measured at wavelength $\lambda$, respectively, $\Shet$ is the spectrum of the
unocculted photosphere determined by $\Thet$, and $\Schord$ is the spectrum of
the portion of the photosphere inside of the transit chord. Following previous
studies (e.g., \citealt{mccullough2014, rackham2017}), we use PHOENIX stellar
spectra \citep{husser2013} to model the emergent spectra of the photospheric
components.

As \citet{rackham2018} note, this formalism assumes that the transit chord can
be described by a single emergent spectrum. Although this is not guaranteed for
any one of our transits, they note that this formalism also holds for transits
in which an occulted spot or faculae crossing event is present in the transit
signal above the observational uncertainty and taken into account in the transit
modeling. We also explored more complex models including multiple spot and
faculae covering fractions but we found that the data did not warrant the
additional complexity of such a model.

Our combined retrieval approach uses the posterior Bayesian evidence $Z \equiv
\mathbb P(D|H)$ computed by \texttt{PyMultiNest}, which is the probability of
the data $D$ given the hypothesis $H$, to perform model comparisons. This
property of the nested sampling algorithm allows us to study how complex our
models have to be to explain the observed distortions to the light curve (such
as number of spots) via the posterior odds,
$\mathbb P(H_n|D) / \mathbb P(H_k|D)$, where the joint probability
$\mathbb P(H_n|D) = P(D|H_n)\mathbb P(H_n)$, with $\mathbb P(H_n)$ being the
prior probability on the hypothesis $H_n$. If we approximate model $n$ and model
$k$ as having the same prior distribution on their respective hypothesis
$H_n, H_k$, then the posterior odds simplify to just the ratio of the evidences,
\begin{align} 
    \frac{\mathbb P(H_n|D)}{\mathbb P(H_k|D)}
    = \frac{\mathbb P(D|H_n)}{\mathbb P(D|H_k)} \equiv \frac{Z_n}{Z_k}\quad.
\end{align}
In log-space, this is the difference of the logs of each evidence and is denoted
as $\Delta \ln Z \equiv \ln Z_n - \ln Z_k$. In the context of this study,
$\Delta\ln Z$ is a measure of how statistically different a given model
($Z_n$) is from a flat atmospheric model ($Z_k \equiv Z_\text{flat}$).
\cite{trotta2008} and \cite{benneke_seager2013} review how these log-odds
translate to frequentist significance hypothesis testing. We note from that work
that absolute log-odds below 1 are usually considered inconclusive, near 2.5 can
be interpreted as moderate evidence, and higher than 5 can be interpreted as
highly significant. It is important to caution though that frequentist
hypothesis testing has only one null hypothesis, whereas proper Bayesian model
comparison considers a range of possible hypotheses, which limits the comparison
with frequentist methods.

\subsection{Retrieval results} \label{sec:retr_results}
\begin{figure*}[htb] 
    \centering
    \includegraphics[width=\textwidth]{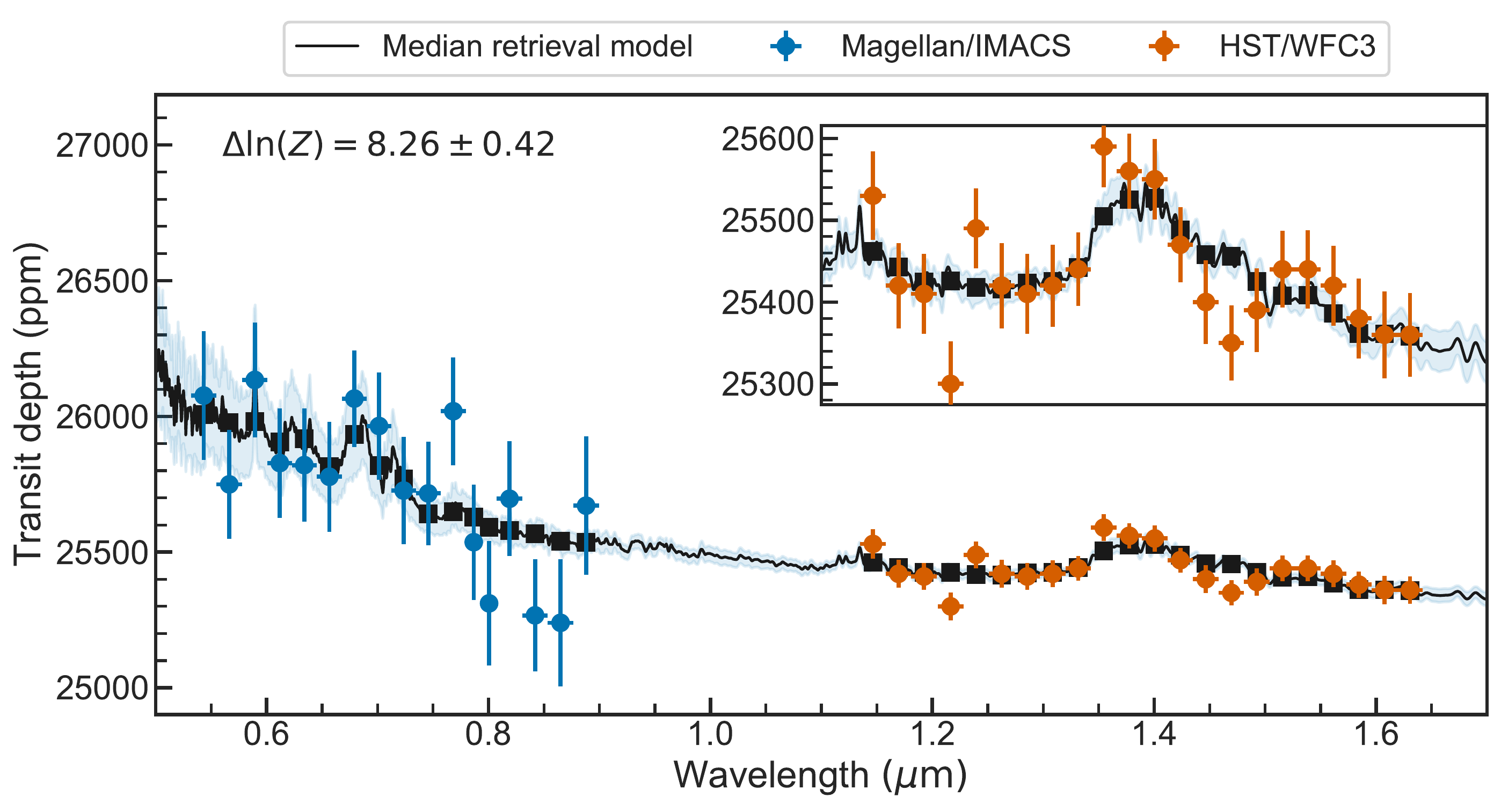}
    \caption{Best retrieved model and 1-sigma uncertainty (highlighted)
    	    for GP+PCA detrended transmission spectrum with data
    	    combined in the visual and NIR. The model includes \water\
    	    in the atmosphere of WASP-43b and stellar heterogeneity. The
    	    black squares show the model binned to this study's
    	    \textit{Magellan/IMACS} data and \textit{HST/WFC3} data
    	    \citep{kreidberg2014}. The top right panel is a zoom-in around the
    	    \textit{HST/WFC3} portion of the spectrum. We share the best
    	    retrieved parameters in Table~\ref{tab:retrieval_table} associated
    	    corner plot in Figure~\ref{fig:corner_retr_GP} of the Appendix.}
    \label{fig:retr_GP}
\end{figure*}

We fit the combined GP+PCA ACCESS detrended spectrum and the HST/WFC3 NIR
spectrum from \cite{kreidberg2014} using the retrieval code described in the
previous section. We fit for an offset between the two datasets, and a range of
models. Those include combinations of clear and cloudy/hazy atmospheres with Na,
K, and \water\ and contamination of the planet's transmission spectrum by
stellar surface heterogeneity. 
Table~\ref{tab:retrieval_priors} shows the prior
distributions of the parameters used in the models. We used the prescription
provided by \citet[Table 2]{benneke_seager2013} to interpret our relative log 
evidences (Table~\ref{tab:lnZ}).
Based on their values, we find that they are all register as a strong detection, 
with no one model being statistically more likely than another. Although, the 
model with the largest log evidence relative to a flat atmosphere 
($\Delta\ln Z = 8.26$) is the one including stellar heterogeneity, combined 
with a clear atmosphere with \water\ (but no Na or K). 
We adopt that model as the one that best fits the data and show it in
Figure~\ref{fig:retr_GP}. From this model, we estimate an average spot 
contrast of $\SI{132}{K} \pm \SI{132}{K}$ and covering fraction of
$0.27^{+0.42}_{-0.16}$.
The corner plot for that model solution is shown in Figure~\ref{fig:corner_retr_GP} 
of the Appendix. The parameters of that model and their
uncertainties are summarized in Table~\ref{tab:retrieval_table}.

\begin{deluxetable*}{l|llll}[htb!]
    \caption{Priors used in retrieval models.}  
    \label{tab:retrieval_priors} 
    \tablehead{\colhead{Model component} & \colhead{Parameter} & \colhead{Units} & \colhead{Description} & \colhead{Prior distribution}} 
    \startdata
        Offset & \text{offset} & ppm
        & Offset between \textit{Magellan/IMACS} and \textit{HST/WFC3} data
        & Normal(Mean Depth, 1000 ppm)\\
        \hline
            Base                & $(R_p/R_s)_0$     & -  	& Radius corresponding to the top of the cloud layer or $\tau \gg 1$    & Uniform(0.8,1.2)\\ 
                                & $P_0$ 	    	& bar 	& Reference pressure at $(R_p/R_s)_0$                                   & Log-uniform($10^{-4}$, $1$) \\ 
                                & $T$               & K     & Average temperature planet atmosphere                                 & Uniform(0, 1500) \\
            \hline 
            Atomic features     & $X$               & -     & Mixing ratio of species X                                             & Log-uniform($10^{-14}$, $1$)\\ 
            \hline 
            Haze                & $a$	            & - 	& Amplitude of the haze cross-section power law                         & Log-uniform($10^{-10}$, $10^{20}$) \\
                                & $\gamma$          & -	    & Index of the haze cross-section power law                             & Uniform(-10, 0) \\ 
            \hline 
            Stellar photosphere & $T_\text{occ}$ 	& K 	& Average temperature of the transit chord	                            & Uniform(4,000, 5,000) \\ 
                                & $T_\text{het}$ 	& K 	& Average temperature of the heterogeneous surface features             & Uniform(4,000, 5,000) \\ 
                                & $F_\text{het}$ 	& -     & Fraction of the unocculted photosphere covered by spots	 	        & Uniform(0, 1) 
    \enddata
\end{deluxetable*}

\begin{deluxetable*}{LlR}[htb]
    \caption{Retrieved parameters for best fit retrieval model shown 
            in Figure~\ref{fig:retr_GP}.}
    \label{tab:retrieval_table}
    \tablehead{\colhead{Parameter} & \colhead{Description} & \colhead{Value}}
    \startdata
        F_\text{het}                        & Spot covering fraction        & 0.27^{+0.42}_{-0.16} \\
        T                                   & Planet temperature (K)        & 352.91^{+206.14}_{-125.08} \\
        T_\text{het}                        & Spot temperature (K)          & 4169.40^{+86.10}_{-92.33} \\
        T_\text{occ}                        & Occulted temperature (K)      & 4300.91^{+94.04}_{-77.16} \\
        \log{\text{H}_2\text O}             & Water volume mixing ratio     & -2.78^{+1.38}_{-1.47} \\
        \log{P_0}                           & Reference pressure (bar)      & -1.67^{+1.08}_{-1.16} \\
        \text{offset}_\text{Magellan/IMACS} & Offset between datasets (ppm) & 25749.40^{+101.54}_{-105.70} \\
        f                                   & Planet radius normalization   & 0.99^{+0.00}_{-0.00} 
    \enddata
\end{deluxetable*}

\begin{deluxetable*}{l|C|C|C}[htb]
    \caption{$\Delta\ln Z$ for various models relative to a flat spectrum for the ACCESS
    GP+PCA detrended data, combined with the \cite{kreidberg2014} data. The largest value of 
    8.26 corresponds to the best retrieved transmission spectrum
    model shown in Figure~\ref{fig:retr_GP}.}
    \label{tab:lnZ}
    \tablehead{\colhead{Model} \vline & \colhead{\water} \vline & 
    \colhead{Na+\water} \vline & \colhead{Na+K+\water}}
    \startdata 
        clear     & 4.98 & 6.42 & 6.08 \\
        haze      & 5.43 & 5.76 & 5.60 \\
        spot      & 8.26 & 7.90 & 7.74 \\
        spot+haze & 6.92 & 6.70 & 6.59
    \enddata
\end{deluxetable*}
 
\section{Summary and Conclusions} \label{sec:conclusion}
We have collected, extracted, and combined transmission spectra of WASP-43b from
\textit{Magellan/IMACS} over four transit epochs spanning the years 2015 to
2018. We combined this with IR data from \textit{HST/WFC3} to create a transmission sprectrum with a total wavelength coverage of \SIrange{5318}{16420}{\angstrom}. 
We analyzed the combined spectrum in a dynamic nested sampling framework with NIR
data from \cite{kreidberg2014}, extending up to \SI{16420}{\angstrom} to search
for the presence of different species. Assuming a water volume mixing ratio 
of $6.1\times10^{-4}$ for a planetary atmosphere with solar abundances 
\citep{kreidberg2014}, our retrieval yields a log \water\ volume 
mixing ratio of $-2.78^{+1.38}_{-1.47}.$ ($2.72^{+65.26}_{-0.09}$ solar). 
Our retrieved water abundance is consistent with the $1\sigma$ range found 
by the joint transmission and emission spectrum analysis in 
\citet[$0.4 - 3.5$ solar]{kreidberg2014} and phase curve analysis in 
\citet{stevenson2014}. Our retrieved planetary temperature
is also consistent with the ranges predicted from the more recent 2.5D theoretical 
phase curve retrievals of the terminator region in \citet[Figure 9]{irwin2019}. 
In the visual spectrum we 
do not observe a statistically significant excess of \ion{K}{1} given the data, 
as reported in \citet{murgas2014}. We also do not observe the presence of
\ion{Na}{1} or \halpha\ in our combined spectra.

Our analysis also investigates the contribution of stellar heterogeneities to
observed transmission spectra, given that WASP-43 is an active star. The best
fitting model to our transmission spectrum calls for the presence of spots in
the surface of the star being a more favored model that a model with atmospheric
hazes. However, we do note that the impact on the spectrum from both hazes and
the contamination from surface stellar heterogeneities can be degenerate, and
given the quality of the current data it is not possible to fully distinguish
between the two.

As the sample of available transmission spectra of different planets increases,
it will become useful to compare their measured spectra and other system
parameters (e.g., stellar irradiation levels, spectral type, metallicity, planet
density) to look for possible correlations. Uniform datasets for a wide range of
planets, such as the ones ACCESS is building, will be crucial in this analysis.

\acknowledgments
The results reported herein benefited from support, collaborations and
information exchange within NASA's Nexus for Exoplanet System Science (NExSS)
research coordination network sponsored by NASA's Science Mission Directorate.
This paper includes data gathered with the 6.5 meter Magellan Telescopes located
at Las Campanas Observatory, Chile.  We thank the staff at the Magellan
Telescopes and Las Campanas Observatory for their ongoing input and support to
make the ACCESS observations presented in this work possible. A.J.\
acknowledges support from FONDECYT project 1171208, and by the Ministry for the
Economy, Development, and Tourism's Programa Iniciativa Cient\'{i}fica Milenio
through grant IC\,120009, awarded to the Millennium Institute of Astrophysics
helped improve the quality of this work.

\software{Astropy \citep{astropy}, corner \citep{corner},
Matplotlib \citep{hunter2007}, NumPy \citep{oliphant2006},
PyMC \citep{pymc}, Multinest \citep{feroz2009}, PyMultiNest \citep{buchner2014},
SciPy \citep{scipy}, SPOTROD \citep{beky2014},
batman \citep{kreidberg2015}, george \citep{george}}

\facilities{Magellan:Baade, Smithsonian Institution High Performance Cluster
(SI/HPC)}

\bibliographystyle{yahapj}
\bibliography{references}

\appendix
\begin{figure*}[htb]
    \centering
    \rotatebox{90}{
    \includegraphics[width=\textwidth]{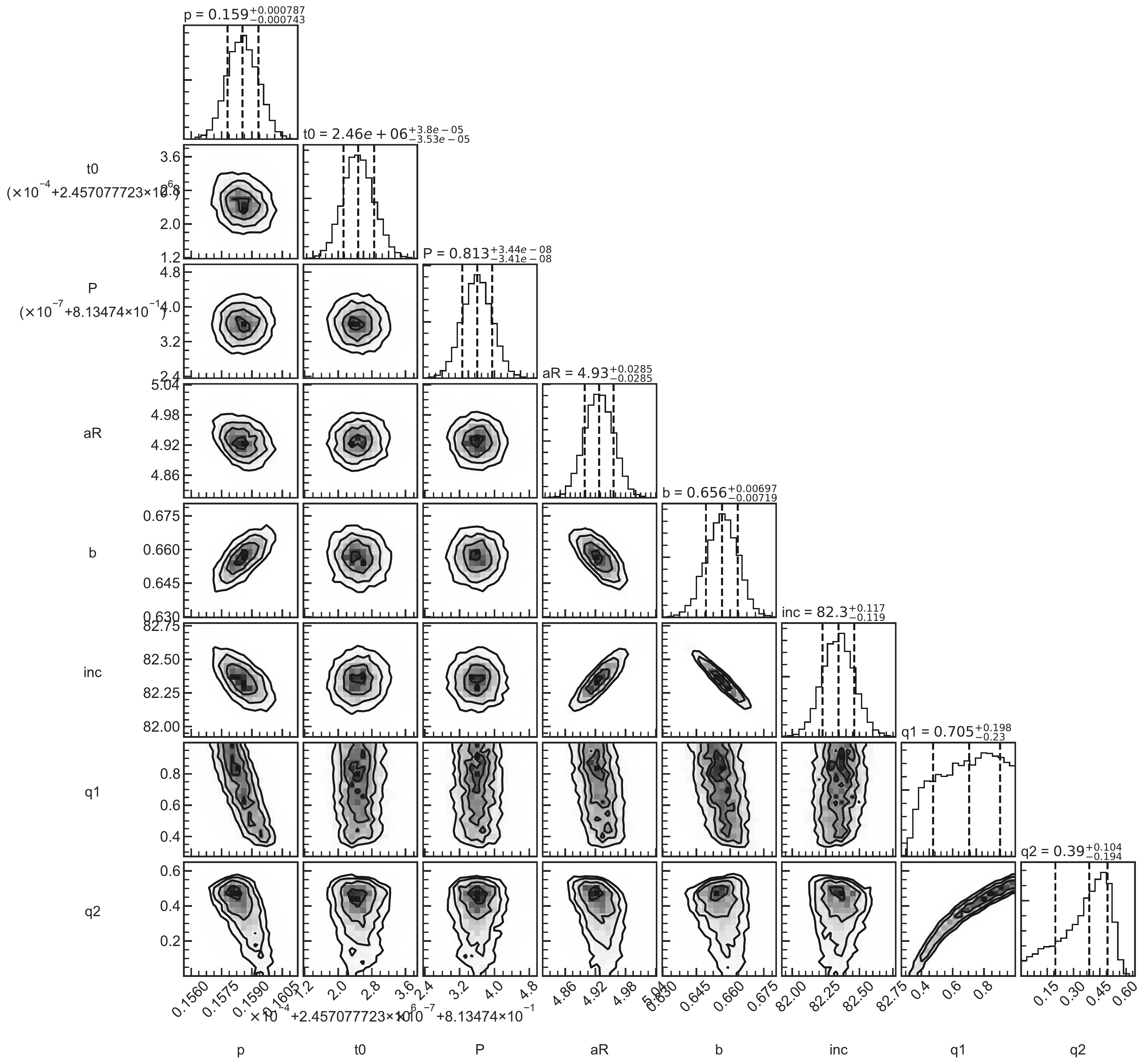}}
    \caption{GP+PCA corner plot of fitted parameters for Transit 1 WLC in
            Figure~\ref{fig:wlc_GP}. Vertical dashed lines mark 16\% and 84\% quantile.
            We share the best fit values in Table~\ref{tab:wlc_GP}.}
    \label{fig:corner_wlc_ut150224_GP}
    \end{figure*} 
\begin{figure*}[htb]
    \centering
    \rotatebox{90}{
    \includegraphics[width=\textwidth]{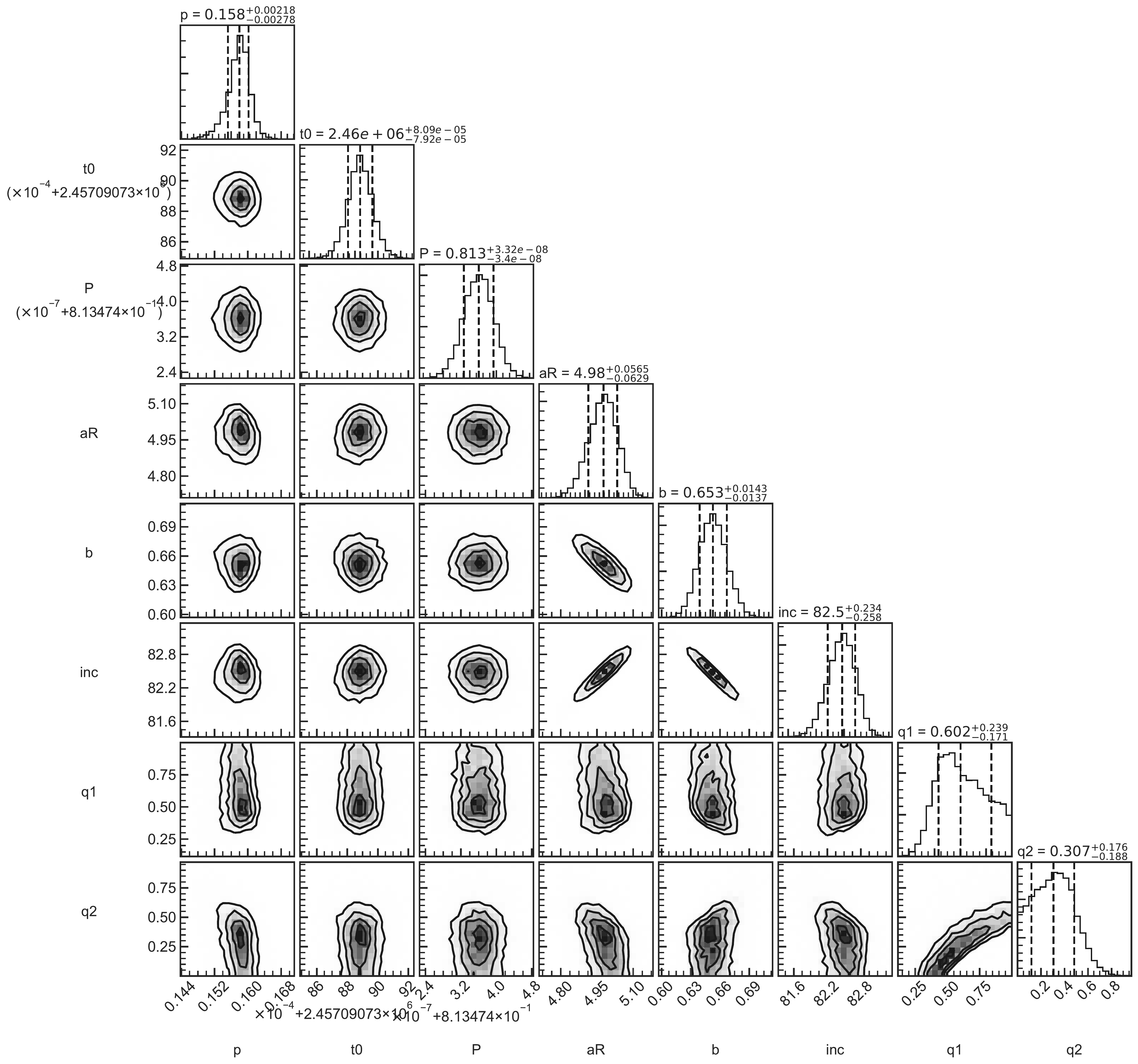}}
    \caption{GP+PCA corner plot of fitted parameters for Transit 2 WLC in
            Figure~\ref{fig:wlc_GP}. Vertical dashed lines mark 16\% and 84\% quantile.
            We share the best fit values in Table~\ref{tab:wlc_GP}.}
    \label{fig:corner_wlc_ut150309_GP}
    \end{figure*} 
\begin{figure*}[htb]
    \centering
    \rotatebox{90}{
    \includegraphics[width=\textwidth]{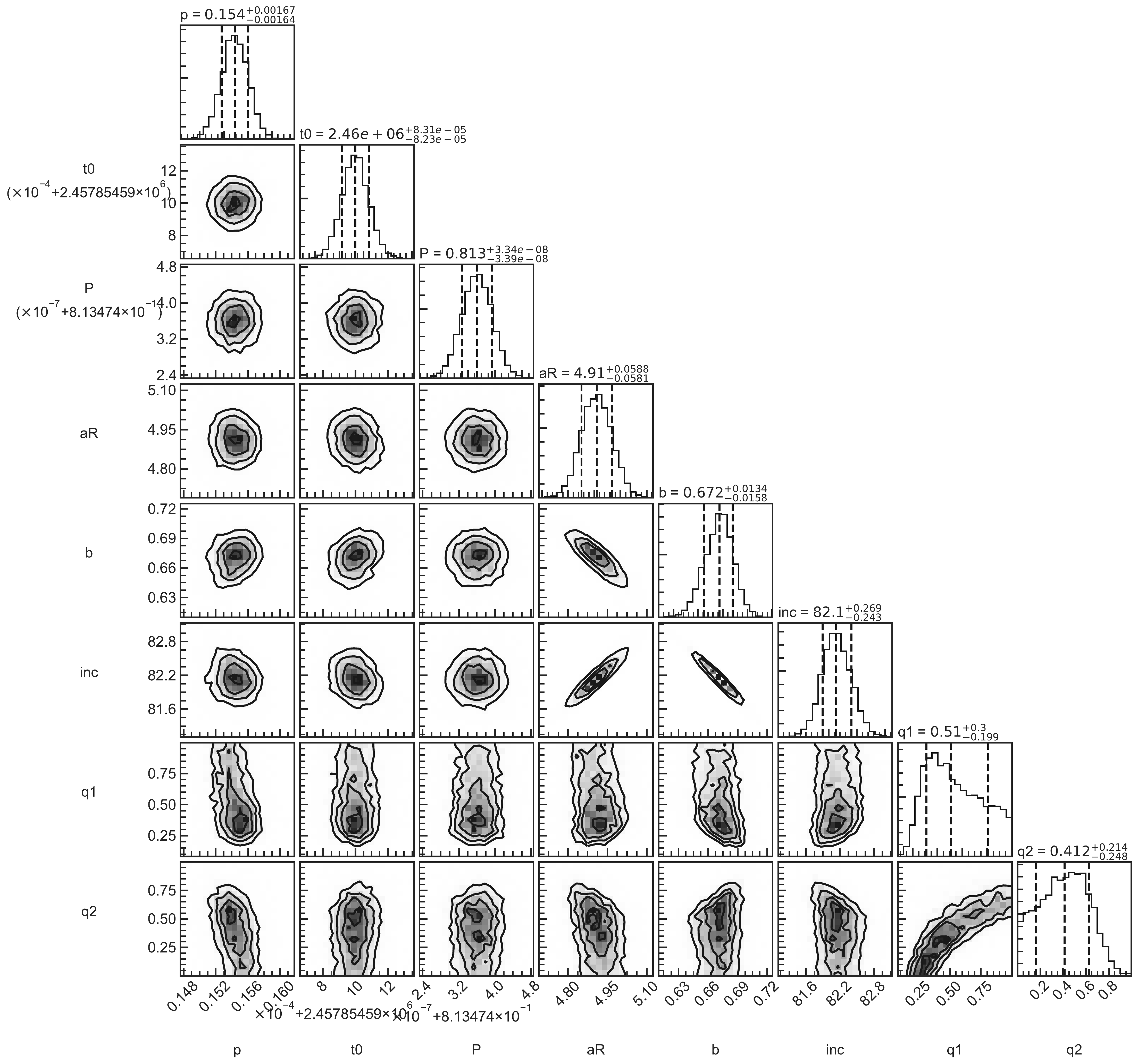}}
    \caption{GP+PCA corner plot of fitted parameters for Transit 3 WLC in
            Figure~\ref{fig:wlc_GP}. Vertical dashed lines mark 16\% and 84\% quantile.
            We share the best fit values in Table~\ref{tab:wlc_GP}.}
    \label{fig:corner_wlc_ut170410_GP}
    \end{figure*} 
\begin{figure*}[htb]
    \centering
    \rotatebox{90}{
    \includegraphics[width=\textwidth]{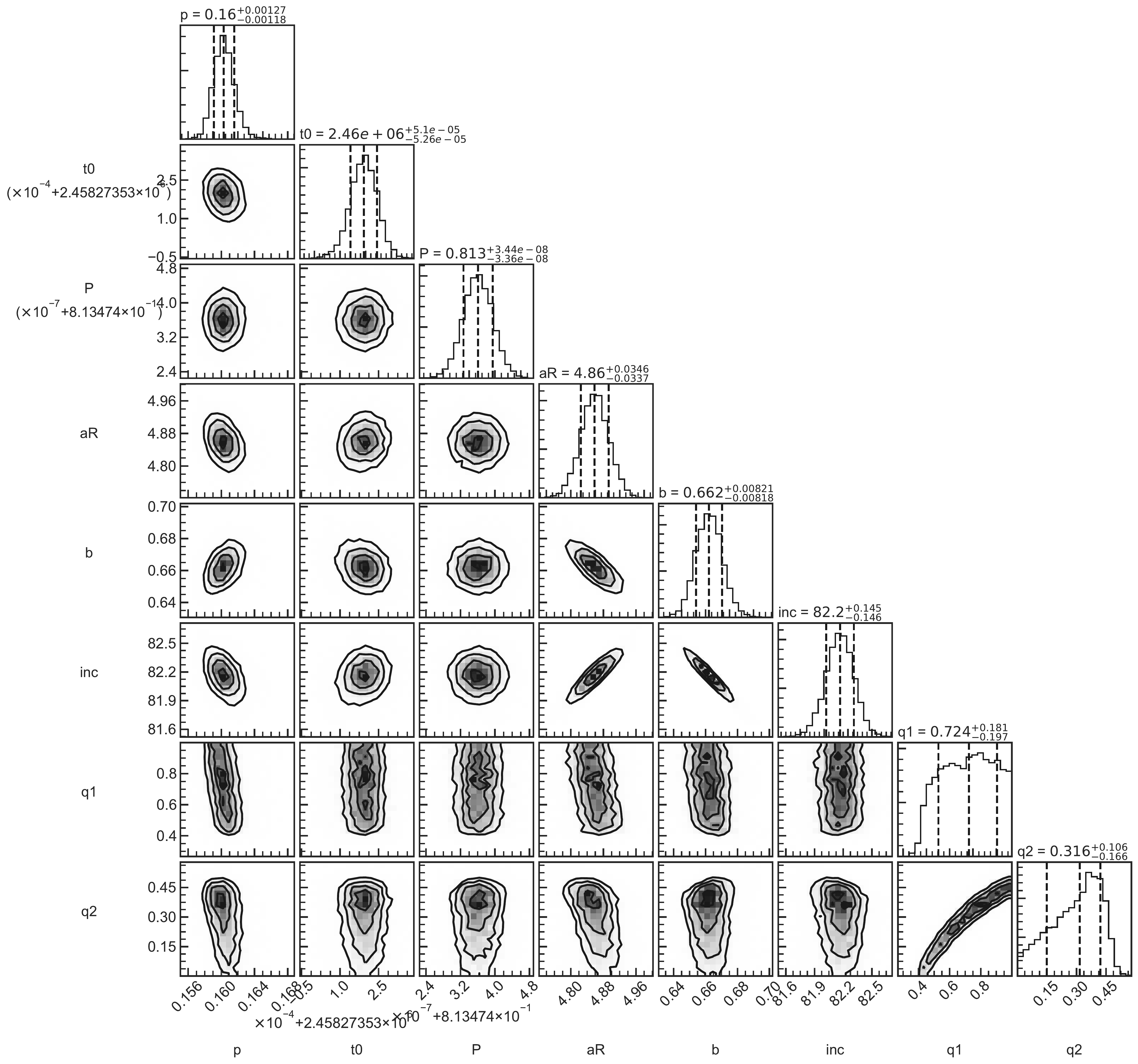}}
    \caption{GP+PCA corner plot of fitted parameters for Transit 4 WLC in
            Figure~\ref{fig:wlc_GP}. Vertical dashed lines mark 16\% and 84\% quantile.
            We share the best fit values in Table~\ref{tab:wlc_GP}.}
    \label{fig:corner_wlc_ut180603_GP}
    \end{figure*} 

\begin{figure}
    \centering
    \rotatebox{90}{
    \includegraphics[width=\textwidth]{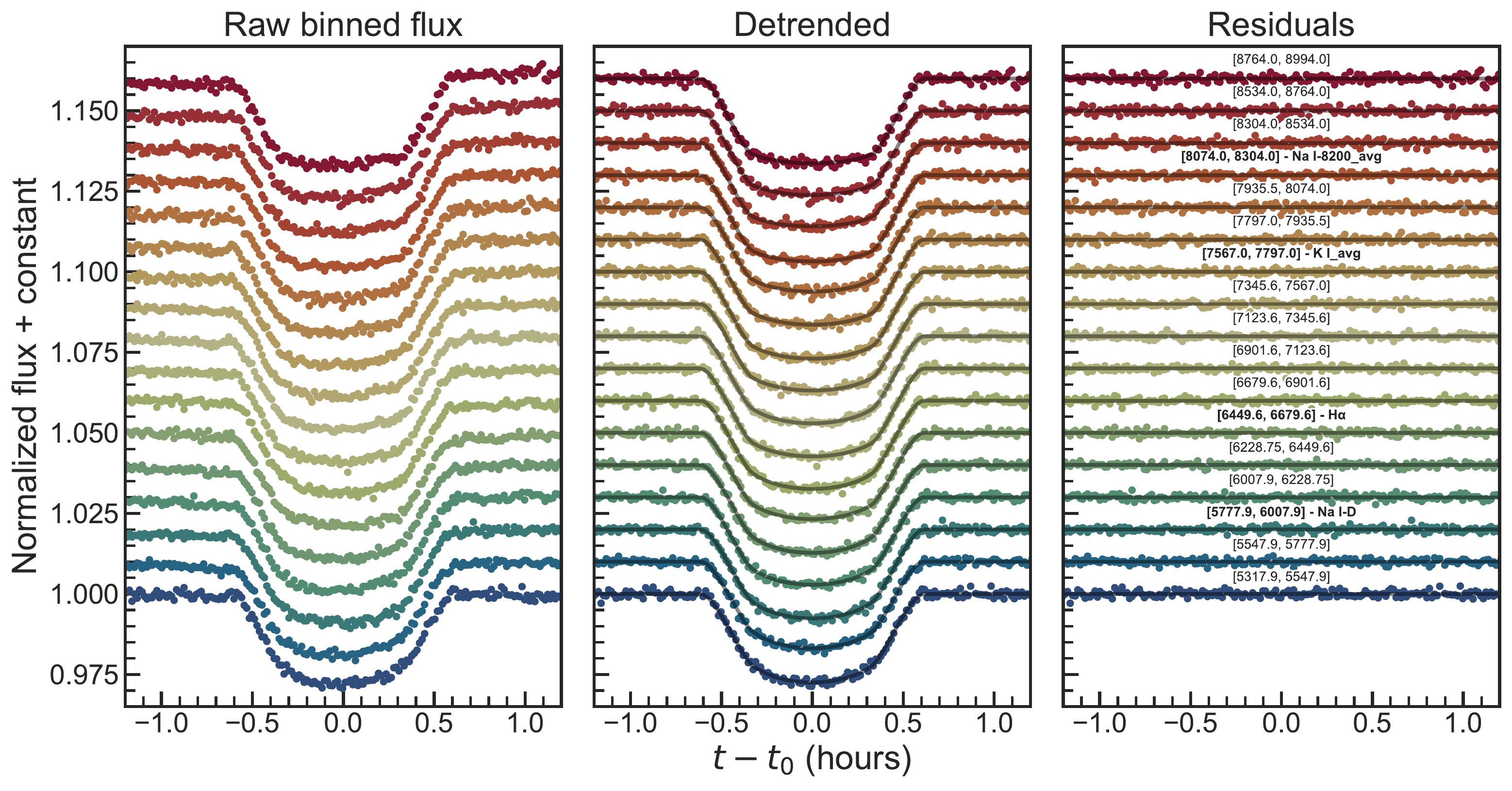}}
    \caption{Binned light curves for Transit 1 shown in 
            Figure~\ref{fig:tspec_combined}. Column 1 shows the raw
    	    observed flux, Column 2 shows the GP+PCA detrended flux and model,
    	    and Column 3 shows the residuals. We labeled the wavelength range of
    	    each bin in Column 3 as well and marked the bins centered around the
    	    vacuum wavelength of potential features of interest in bold. We
    	    centered all data 1 hour around the fitted mid-transit time $t_0$
    	    from the corresponding WLC in Figure~\ref{fig:wlc_GP}.
    	    We share the binned transit data in Table~\ref{tab:binned_table_gp}.}
    \label{fig:binned_ut150224_GP}
\end{figure}
\begin{figure}
    \centering
    \rotatebox{90}{
    \includegraphics[width=\textwidth]{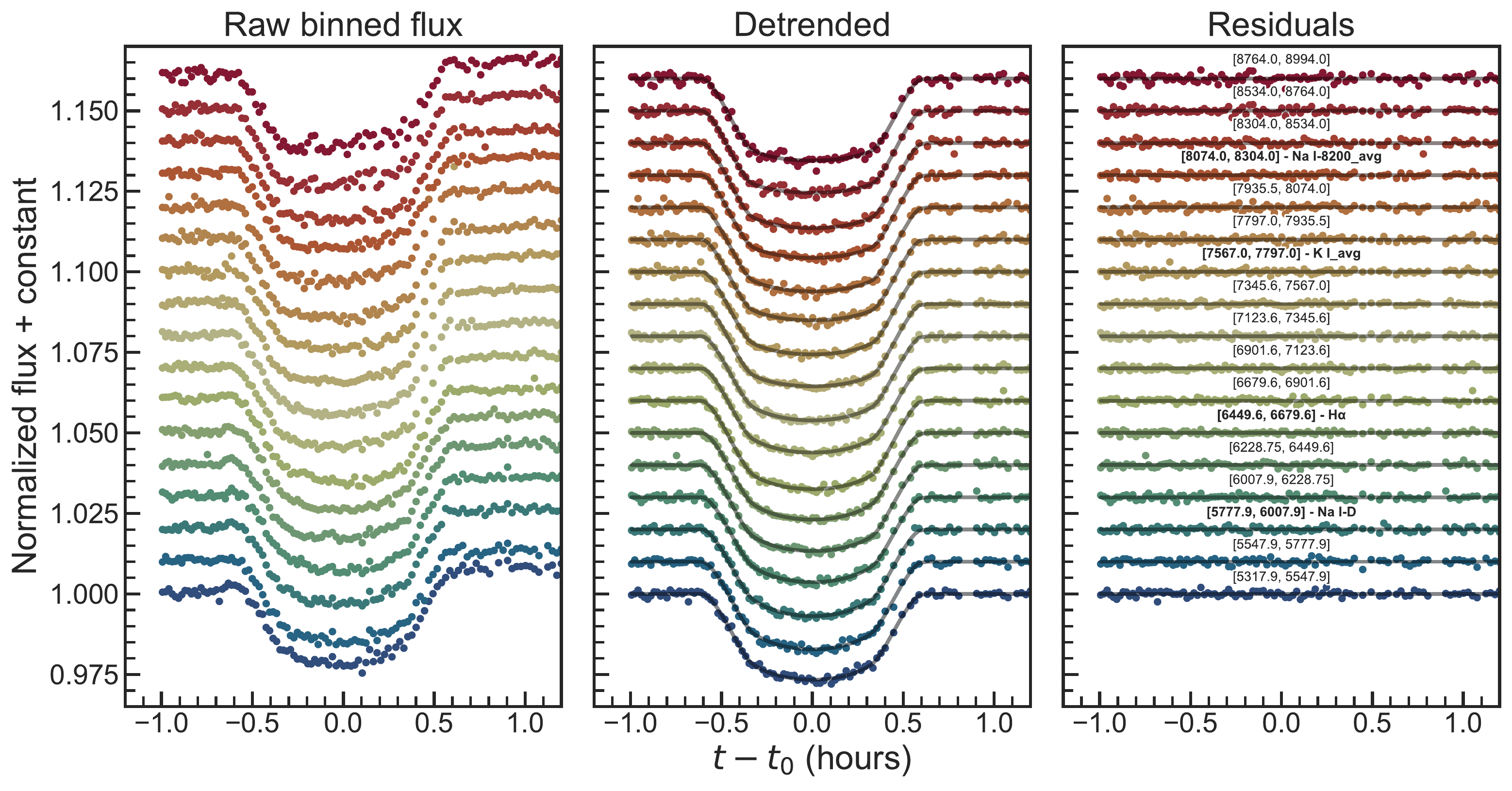}}
    \caption{Binned light curves for Transit 2 shown in
            Figure~\ref{fig:tspec_combined}. Column 1 shows the raw
    	    observed flux, Column 2 shows the GP+PCA detrended flux and model,
    	    and Column 3 shows the residuals. We labeled the wavelength range of
    	    each bin in Column 3 as well and marked the bins centered around the
    	    vacuum wavelength of potential features of interest in bold. We
    	    centered all data 1 hour around the fitted mid-transit time $t_0$
    	    from the corresponding WLC in Figure~\ref{fig:wlc_GP}.
    	    We share the binned transit data in Table~\ref{tab:binned_table_gp}.}
    \label{fig:binned_ut150309_GP}
\end{figure}
\begin{figure}
    \centering
    \rotatebox{90}{
    \includegraphics[width=\textwidth]{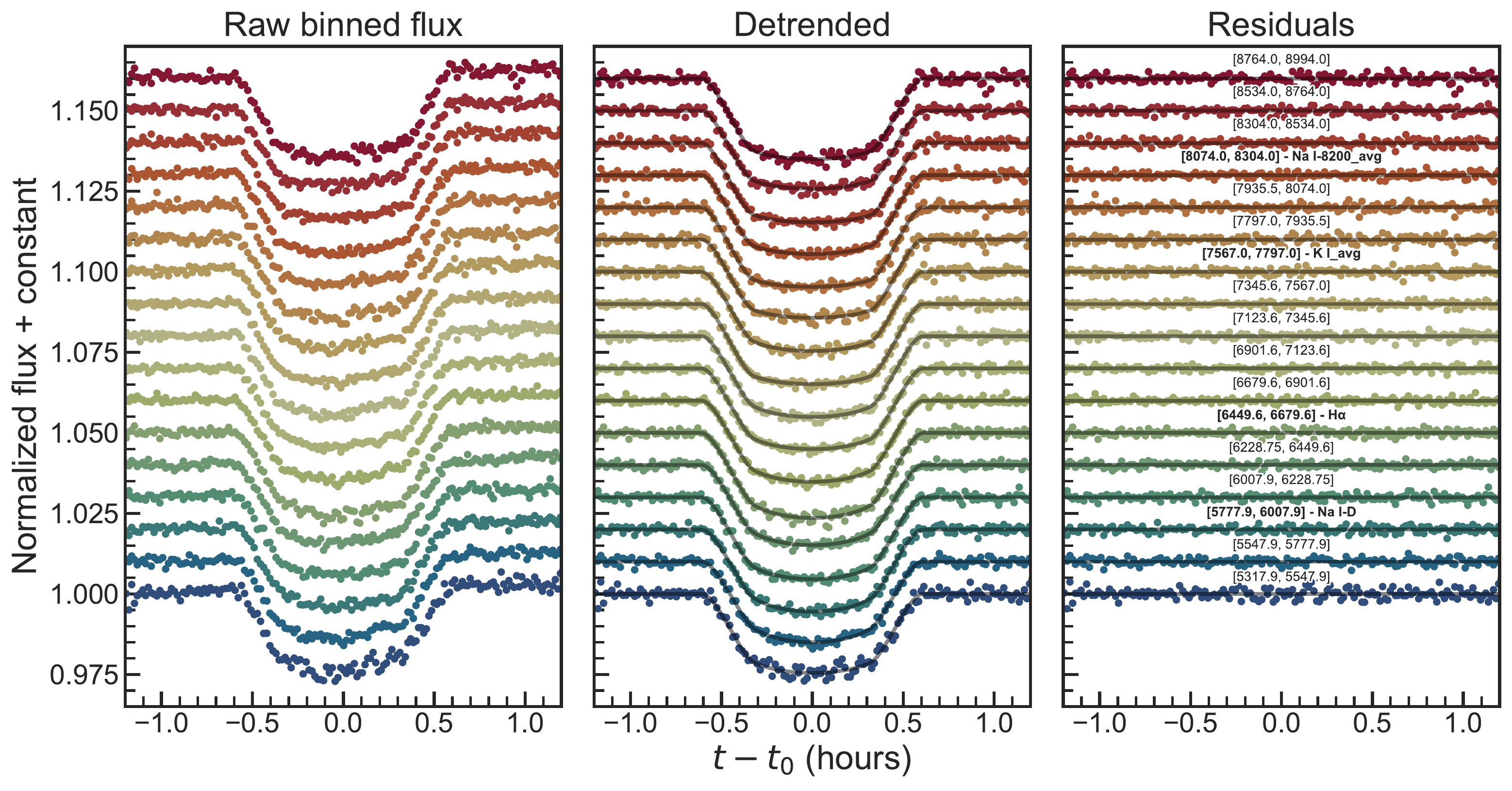}}
    \caption{Binned light curves for Transit 3 shown in
            Figure~\ref{fig:tspec_combined}. Column 1 shows the raw
    	    observed flux, Column 2 shows the GP+PCA detrended flux and model,
    	    and Column 3 shows the residuals. We labeled the wavelength range of
    	    each bin in Column 3 as well and marked the bins centered around the
    	    vacuum wavelength of potential features of interest in bold. We
    	    centered all data 1 hour around the fitted mid-transit time $t_0$
    	    from the corresponding WLC in Figure~\ref{fig:wlc_GP}.
    	    We share the binned transit data in Table~\ref{tab:binned_table_gp}.}
    \label{fig:binned_ut170410_GP}
\end{figure}
\begin{figure}
    \centering
    \rotatebox{90}{
    \includegraphics[width=\textwidth]{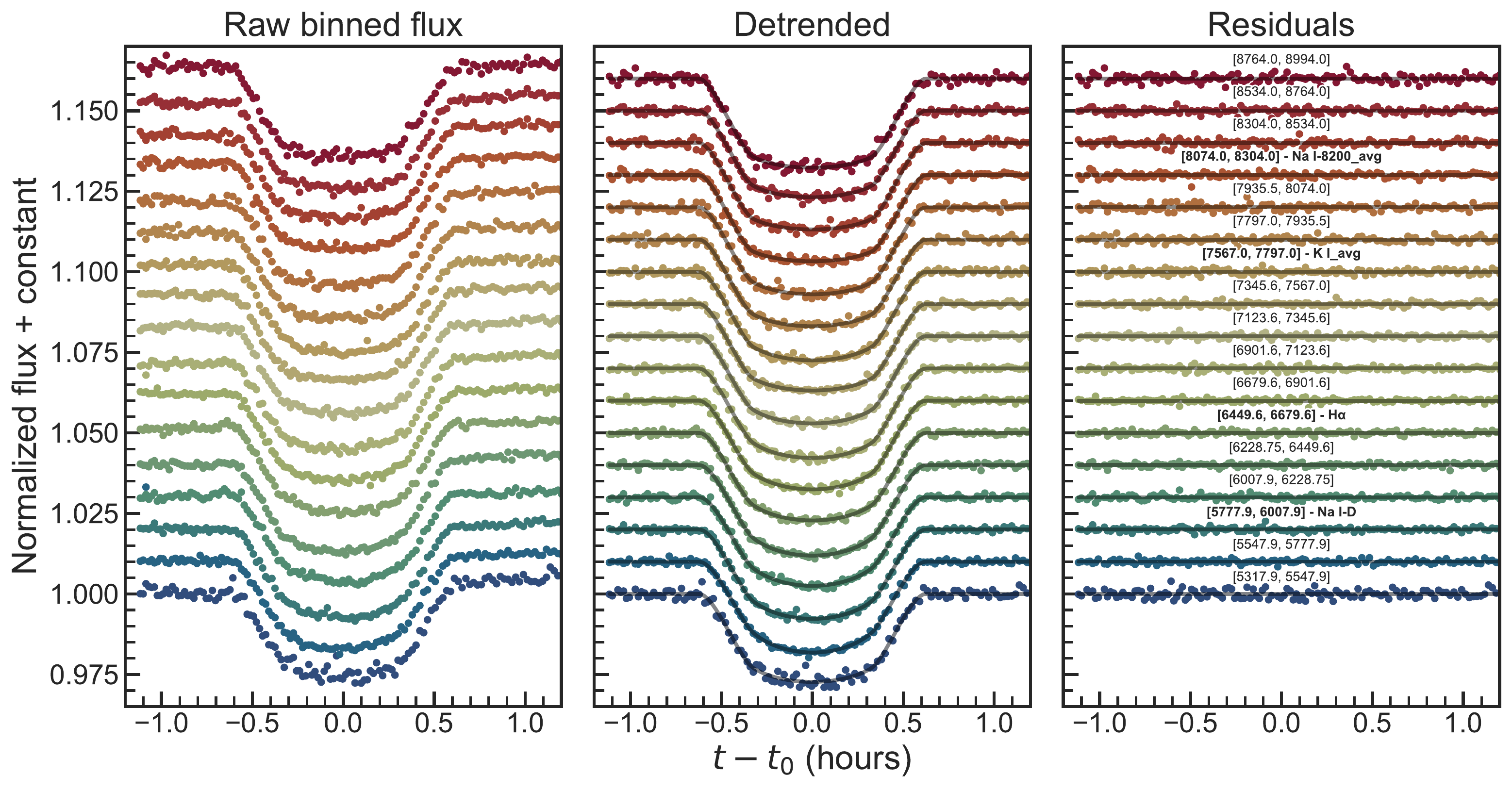}}
    \caption{Binned light curves for Transit 4 shown in
            Figure~\ref{fig:tspec_combined}. Column 1 shows the raw
    	    observed flux, Column 2 shows the GP+PCA detrended flux and model,
    	    and Column 3 shows the residuals. We labeled the wavelength range of
    	    each bin in Column 3 as well and marked the bins centered around the
    	    vacuum wavelength of potential features of interest in bold. We
    	    centered all data 1 hour around the fitted mid-transit time $t_0$
    	    from the corresponding WLC in Figure~\ref{fig:wlc_GP}.
    	    We share the binned transit data in Table~\ref{tab:binned_table_gp}.}
    \label{fig:binned_ut180603_GP}
\end{figure}
    
\begin{deluxetable*}{CCCCCC}[htb]
    \caption{Associated transit depths of Figure~\ref{fig:tspec_combined}, 
    shown relative to the mean fitted GP+PCA WLC depth of \SI{25071}{ppm}.}
    \label{tab:binned_table_gp}
    \tablehead{\colhead{Wavelength (\angstrom) } & \colhead{Transit 1} & \colhead{Transit 2} & \colhead{Transit 3} & \colhead{Transit 4} & \colhead{combined}} 
    \startdata
    5317.9 - 5547.9  & 523.0603 \pm 260.4245  & -267.3194 \pm 938.3293  & -1922.6752 \pm 1020.5674 & 522.3093 \pm 1102.5989 & 339.2619 \pm 237.9392  \\
    5547.9 - 5777.9  & -116.1220 \pm 266.1869 & 172.6987 \pm 589.2437   & -1455.4281 \pm 596.1788  & 1073.0221 \pm 440.0050 & 12.2549 \pm 200.1120   \\
    5777.9 - 6007.9  & 589.0840 \pm 278.1240  & -148.8073 \pm 761.4993  & -1069.2411 \pm 595.0667  & 936.0414 \pm 455.4287  & 397.0056 \pm 211.7735  \\
    6007.9 - 6228.75 & 295.5726 \pm 269.8283  & -542.7774 \pm 715.1426  & -1145.5722 \pm 635.7125  & 320.1664 \pm 392.9886  & 90.7797 \pm 201.4572   \\
    6228.75 - 6449.6 & 246.0972 \pm 273.4700  & -250.3903 \pm 740.0324  & -1642.9432 \pm 528.5344  & 1157.7405 \pm 482.7322 & 83.2450 \pm 208.2044   \\
    6449.6 - 6679.6  & -79.6483 \pm 287.1381  & 24.9089 \pm 660.5415    & 35.8304 \pm 435.9645     & 373.9576 \pm 469.4392  & 40.3179 \pm 203.1956   \\
    6679.6 - 6901.6  & 495.3958 \pm 236.3006  & 886.1126 \pm 651.1368   & -1194.4574 \pm 482.4106  & 651.8302 \pm 380.5127  & 328.2793 \pm 178.2558  \\
    6901.6 - 7123.6  & 632.9288 \pm 254.1061  & -785.1288 \pm 655.7699  & -1535.3495 \pm 493.4588  & 1103.0682 \pm 514.3652 & 227.2646 \pm 197.2615  \\
    7123.6 - 7345.6  & 300.9717 \pm 245.2279  & -700.5517 \pm 718.5057  & -1524.4228 \pm 551.7447  & 281.0918 \pm 508.2007  & -10.5521 \pm 197.1705  \\
    7345.6 - 7567.0  & 291.1467 \pm 263.7409  & -974.6394 \pm 713.1503  & -1460.5078 \pm 485.1884  & 498.0201 \pm 384.7823  & -20.8910 \pm 191.2337  \\
    7567.0 - 7797.0  & 493.7086 \pm 254.0115  & -726.8098 \pm 960.7674  & -1840.4984 \pm 616.9723  & 832.4674 \pm 403.5181  & 281.7274 \pm 198.6125  \\
    7797.0 - 7935.5  & -100.4397 \pm 310.0690 & -1304.6125 \pm 934.2026 & -1739.2184 \pm 566.2540  & 479.8308 \pm 368.1740  & -200.6929 \pm 212.9927 \\
    7935.5 - 8074.0  & -453.8738 \pm 302.4915 & -372.3211 \pm 954.3857  & -1253.9094 \pm 537.2197  & 463.4859 \pm 534.4578  & -426.0211 \pm 229.4610 \\
    8074.0 - 8304.0  & 277.2830 \pm 287.8497  & -728.4844 \pm 809.5412  & -1631.6495 \pm 572.5958  & 318.7489 \pm 417.2814  & -40.6945 \pm 211.3458  \\
    8304.0 - 8534.0  & -593.7829 \pm 265.2407 & -134.9286 \pm 843.3740  & -1441.1786 \pm 556.1943  & 460.0621 \pm 458.6014  & -470.2350 \pm 205.8146 \\
    8534.0 - 8764.0  & -100.5637 \pm 328.7061 & -896.6887 \pm 954.9673  & -2019.0232 \pm 479.6925  & 449.1528 \pm 529.7768  & -497.6864 \pm 234.0147 \\
    8764.0 - 8994.0  & -3.1396 \pm 354.2233   & -822.3481 \pm 1178.9426 & -1235.4496 \pm 531.3372  & 1244.3340 \pm 559.6876 & -66.1594 \pm 254.6280
    \enddata 
\end{deluxetable*}

\begin{figure*}
    \centering
    \rotatebox{90}{
    \includegraphics[width=\textwidth]{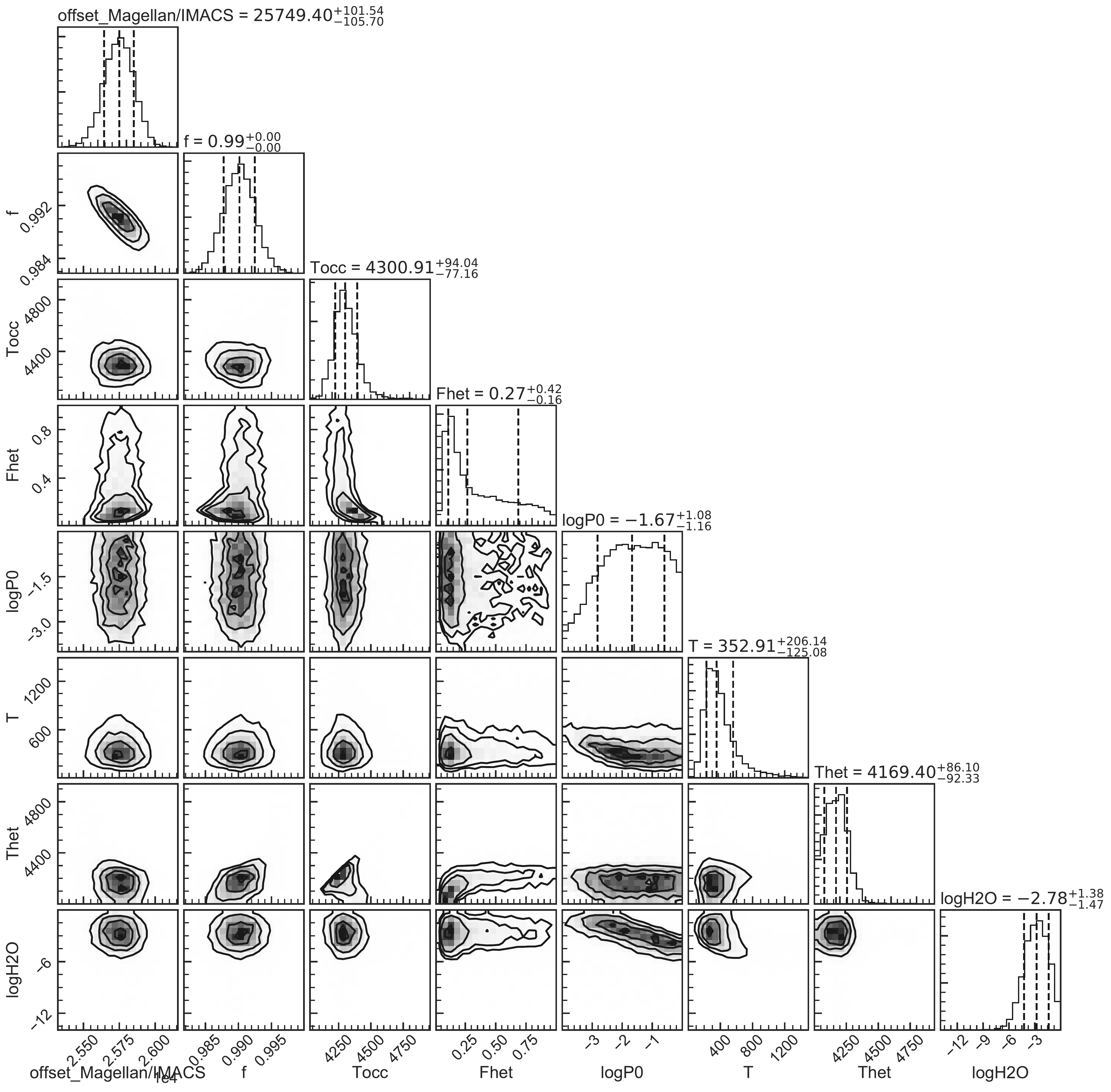}
    }
    \caption{Corner plot for the best fit transmission spectrum
            retrieved in Figure~\ref{fig:retr_GP}.}
    \label{fig:corner_retr_GP} 
\end{figure*}

\end{document}